\documentclass[review,authoryear,12pt]{elsarticle}
\usepackage{graphicx}
\usepackage{amssymb} 
\usepackage{amsmath} 

\journal{Ecological Modelling}

\begin{document}

\begin{frontmatter}

\title{Exploring the tug of war between positive and negative interactions among savanna trees:
Competition, dispersal, and protection from fire.}

\author[1,4]{Flora S. Bacelar \corref{cor1}}\ead{fbacelar@ufba.br}
\cortext[cor1]{Corresponding author:  Tel:~+55-71-32836600; fax:~+55-71-3283-6606.}
\author[2,3]{Justin M. Calabrese}
\author[1]{Emilio Hern\'{a}ndez-Garc\'{\i}a}
\address[1]{IFISC(CSIC-UIB) Instituto de F{\'\i}sica Interdisciplinar y Sistemas
Complejos, Campus Universitat de les Illes Balears, E-07122
Palma de Mallorca, Spain}
\address[2]{Conservation Ecology Center, Smithsonian Conservation Biology Institute,
National Zoological Park, 1500 Remount Rd., Front Royal, VA 22630, USA}
\address[3]{Department of Ecological Modelling, Helmholtz Centre for Environmental
Research--UFZ, Permoserstrasse 15, 04318 Leipzig, Germany}
\address[4] {Instituto de F\'{\i}sica,
Universidade Federal da Bahia, 40210-340, Salvador-Bahia, Brasil}

\begin{abstract}
Savannas are characterized by a discontinuous tree layer
superimposed on a continuous layer of grass. Identifying the
mechanisms that facilitate this tree-grass coexistence has
remained a persistent challenge in ecology and is known as the
``savanna problem". In this work, we propose a model that
combines a previous savanna model (Calabrese et al., 2010),
which includes competitive interactions among trees and
dispersal, with the Drossel-Schwabl forest fire model,
therefore representing fire in a spatially explicit manner. The
model is used to explore how the pattern of fire-spread,
coupled with an explicit, fire-vulnerable tree life stage,
affects tree density and spatial pattern. Tree density depends
strongly on both fire frequency and tree-tree competition
although the fire frequency, which induces indirect
interactions between trees and between trees and grass, appears
to be the crucial factor controlling the tree-extinction
transition in which the savanna becomes grassland. Depending on
parameters, adult trees may arrange in different regular or
clumped patterns, the later of two different types (compact or
open). Cluster-size distributions have fat tails but clean
power-law behavior is only attained in specific cases.
\end{abstract}


\date{\today}

\end{frontmatter}


\section{Introduction}

Savanna ecosystems are characterized by the robust coexistence
of trees and grass. The mechanisms allowing for the persistence
of both types of vegetation and governing the population
dynamics and spatial arrangement of savanna trees are poorly
understood \citep{scholes_tree-grass_1997, bond_what_2008}. Of
the many potential driving mechanisms investigated, local-scale
interactions among trees have received increasing attention in
recent years \citep{barot_demography_1999,
wiegand_patch-dynamics_2006, meyer_multi-proxy_2008,
meyer_rhythm_2007, meyer_satchmo_2007, scanlon2007,
Calabrese2010}. Such tree-tree interactions can roughly be
divided into two classes: facilitative and competitive.
Facilitation among trees promotes tree clustering and may be
mediated by a variety of mechanisms such as limited-range
dispersal, improvement of local resource conditions, and
protection from fire \citep{belsky_effects_1989, Hochberg1994,
holdo_stem_2005, scanlon2007, Calabrese2010}.

Alternatively, competition among trees for water, nutrients,
and light may constrain tree density and favor tree-grass
coexistence, as well as promoting spatial separation between
trees \citep{barot_demography_1999, meyer_multi-proxy_2008,
Calabrese2010}.

There is evidence for both classes of interactions in the
savanna literature, sometimes coming from the same region. For
example, several studies have found evidence consistent with
competition in the Kalahari \citep{skarpe_spatial_1991,
jeltsch_detecting_1999, moustakas_long-term_2006,
moustakas_spacing_2008, meyer_multi-proxy_2008}, while others
have found evidence suggesting facilitation
\citep{caylor_tree_2003, scanlon2007}. Indeed, one of the key
difficulties in understanding the forces structuring savanna
tree populations is that both classes of local-scale
interactions often occur together and it is not obvious whether
the net effect of local interactions on tree population
dynamics will be positive or negative \citep{bond_what_2008}.
Further studies, both empirical and theoretical, are needed to
better understand the interplay between these opposing forces.
Specifically, studies that focus on a limited number of
processes and their interactions should help illuminate the
conditions under which positive or negative local interactions
structure savanna tree populations.

Mesic savannas that receive ~400-800 mm of mean annual
precipitation (MAP) are particularly interesting because there
is evidence from such systems that, in addition to local-scale
interactions, fire plays an important role.
\citep{sankaran_determinants_2005, bucini_continental_2007}.
Both of these factors can act strongly on juvenile trees and
can contribute to a demographic bottleneck through which
juvenile trees must pass to recruit into the adult population.
In contrast to forest tree species, savanna trees are often
more fire resistant \citep{Hoffmann2003}, thus savanna fires
effectively burn the grass layer and the young trees included
in it, leaving adult trees alive, affecting only tree
recruitment and not adult survival \citep{Gignoux1997}. Recent
studies highlighting the importance of tree competition and/or
fire on savannas are \citet{higgins_fire_2000},
\citet{moustakas_long-term_2006},
\citet{moustakas_spacing_2008}, \citet{Odorico2006},
\citet{Hanan2008}, \cite{meyer_multi-proxy_2008}, or
\citet{Calabrese2010}. From their results we might expect a
kind of tug of war between these forces, the outcome of which
affects both the tree-grass balance of the savanna and the
spatial arrangement of adult trees.

The role of fire in mesic savannas is two-fold. On the one
hand, it provides an indirect way for grass to compete against
trees: the higher recovery rates of grasses compared to
juvenile trees make grass the dominant form of vegetation
shortly after a fire has destroyed both. On the other hand,
several studies have suggested that adult trees can protect
vulnerable juveniles from fire, thus increasing their chances
of survival \citep{Hochberg1994, holdo_stem_2005}, but this protection effect has not been intensively studied.
However, given the frequent occurrence of fires in many
savannas, it seems likely that the protection effect may be one
of the most common facilitative interactions among savanna
trees, and the dominance of grass after fire could be as
important as tree-tree competition in restricting the amount of
tree-cover in the savanna.

Recently, \citet{Calabrese2010} studied the interaction between
competition and fire in a highly simplified savanna model. They
showed that these two forces interact non-linearly with
sometimes surprising consequences for tree population density
and spatial pattern. However, because \citet{Calabrese2010}
treated fire in a non-spatially explicit manner, only the
negative impact on trees, and not the protection effect, was
included and thus they could not fully tease apart how these
contrasting local interactions function in combination.

Here, we focus on a spatially explicit lattice model of savanna
tree and grass population dynamics under the influence of
competition and fire. The model is an extension of the
semi-spatial model studied by \citet{Calabrese2010}.
Importantly, both competition and fire are spatially explicit
processes in the new model. This allows us to study directly
how adult trees influence the survival probabilities of nearby
juveniles. We treat competition in the same way as in
\citet{Calabrese2010} and fire is implemented in a similar
manner as in the Drossel-Schwabl forest fire model from
statistical physics \citep{Drossel1992}. In contrast to adult trees in the Drossel-Schwabl model, grasses and juvenile trees are the
flammable objects in our case. We highlight the ranges of
conditions under which local interactions result in net
positive and net negative influences on juvenile tree
recruitment, and we demonstrate how these local interactions
affect the density and spatial structure of adult-tree
populations.

\section{Spatially explicit fire models}
\label{sec:firemodels}

\citet{Bak1990} introduced a simple forest fire model to
demonstrate the emergence of scaling and fractal energy
dissipation. \citet{Drossel1992} extended this model by
introducing a lightning or sparking parameter $f$, and this is
the forest fire model we have adapted to study fire spread in
savannas. It is one of the best studied examples of
non-conservative, self-organized criticality
\citep{Bak1990,Grassberger1991,Drossel1992,Clar1996,Clar1999,schenk2000}.
The forest fire model is a probabilistic cellular automaton
defined on a 2-dimensional lattice of $L^2$ sites, initialized
with a combination of burning trees and live trees, and updated
at each time-step with the following four simple rules: (i) A
burning tree becomes an empty site. (ii) A live tree becomes a
burning tree if at least one of its nearest neighbors is
burning. Some immunity can be introduced in this rule, so that
a green tree becomes a burning tree with probability $1-I$
\citep{Clar1996}. (iii) A new tree establishes at an empty site
with probability $r$. (iv) Live trees in the lattice
spontaneously (i.e., without the need of a burning neighbor)
ignite with probability $f$. This model displays very rich
behavior, and depending on the parameters $f$ and $r$, it
features spiral-like fronts, critical states and phase
transitions. Furthermore, while the Drosel-Schwabl model is
minimalistic, it produces burn patterns similar to those
observed empirically, and is closely related to more detailed
wildfire models \citep{Zinck2009}.

\section{Savanna Fire Model (SFM)}
\label{sec:SFM}

Our model is run in a square lattice with periodic boundary
conditions. We use a lateral size of $L=200$ sites, so that
there are $N=L\times L=4\times10^4$ lattice sites in the
simulation domain. Each site represents a savanna square of $5$
meters on a side. In the previous savanna model (SM) of
\citet{Calabrese2010}, each site in the lattice could be in one
of two states: grass- or tree-occupied. The savanna fire model
(SFM) introduced here combines the previous SM and the
above-described Drossel-Schwabl Forest Fire model, but with the
flammable components being grass and juvenile trees. In this
way, fire is included explicitly as a possible state in the
dynamics. The SFM considers three new states in addition to the
two in the SM so that each site on the lattice can be in one of
the following five states: Grass (G), Juvenile Tree (JT), Adult
Tree (AT), Burning (B) and Ashes (A).

We can distinguish two interaction neighborhoods for each
lattice site: the {\sl near} neighborhood consists of the eight
sites sharing an edge or a corner with the central one (Moore
neighborhood), and we assume this is the spatial scale at which
direct competition among trees occurs. The {\sl far}
neighborhood consists of the sixteen additional sites
surrounding the {\sl near} ones and sharing edges or corners
with them. They will be assumed to be the farthest sites to
which seeds from a focal tree can arrive.

We note that fire propagation occurs over a much shorter
timescale (the spread rate may be around 2 m/s, see
\cite{Cheney1995}) than tree growth, reproduction, death, and
other ecological processes. Thus we implement the burning
process on top of the previous SM, but acting on a faster
scale. Specifically, at each time step, time advances by
$\Delta t=0.1$ years, and the whole lattice is scanned in
parallel to check for one of the following updates:

\begin{enumerate}

   \item Growth: A random number is drawn for each site
       occupied by a juvenile tree so that with probability
       $m \Delta t$ it becomes an adult tree. Thus $m^{-1}$
       is the mean time for a juvenile tree to become
       adult.

   \item Reproduction and establishment:  Each adult tree
       in the lattice sends, with probability $b \Delta
       t/24$, a seed to each of the 24 sites within
       its {\sl near} and {\sl far} neighborhood . If the seed lands on a site in a
       state which is neither G nor A, then nothing happens
       (establishment fails). If instead a site occupied by
       grass or ashes is reached a juvenile tree is
       established.

    \item  Competition: A juvenile tree survives
        competition with neighboring adult trees with
        probability $P_C^{Surv}$. This survival probability
        depends only on the competition exerted by
       neighboring adult trees: $P_C^{Surv}=e^{-\delta S_1}$,
       where $\delta$ is the competition parameter and
       $S_1$ is the number of adult trees in the {\sl near}
       neighborhood.

    \item Death: A random number is drawn for each site
        occupied by an adult tree, so that with probability
        $\alpha \Delta t$ this tree dies. Thus
        $\alpha^{-1}$ is the average adult-tree lifespan.

    \item Recovery: At each time step, each ash site may
        recover into grass with probability $r \Delta t$,
        so that $r^{-1}$ is the mean recovery time of grass
        from ashes. Note that this forces a delay between
        successive fire fronts, thus preventing the lattice from
        continually burning.

  \item Spontaneous burning: There is a ``lightning
      parameter", $f$, so that fire appears spontaneously
      on the lattice at this rate, affecting grass and
      juvenile trees. More explicitly, lattice sites
      occupied by G and JT are checked so that with
      probability $f\Delta t/N$ they become burning sites.

 \item Fire propagation and extinction: After updating with
     all the above processes, a new pass through the
     lattice is done, so that if some fire has been
     introduced in the previous step, fire propagation is
     simulated until the fire burns out. As previously
     mentioned, we assume that this process is fast and
     occurs on a much shorter timescale than the $\Delta
     t=0.1$ years introduced above. It is implemented in
     the following way:

 \begin{description}

\item[a)] Each G and JT site is checked and if at least
    one site in its {\sl near} neighborhood is in the B
    state, the site also burns with probability $1-I$,
    where $I \in [0, 1]$ is an {\sl immunity}
    parameter. This models fire propagation on grass
    and juvenile trees. Note that, since adult trees do
    not burn, fire has a lower chance of reaching JT
    (and G) sites which are surrounded by some adult
    trees. The inclusion of fire in a spatially
    explicit manner therefore implements the protection
    effect from adult trees.

 \item[b)] End of burning:  All sites that were burning
     before entering the previous step (a) are set to
     ashes.

\end{description}

Processes a) and b) are repeated until no burning sites
remain. Time then advances $\Delta t$ units and the
algorithm repeats again from step (1) on the updated
lattice.

 \end{enumerate}

\begin{table}[ht]
\caption{Parameters}
\label{parnamesComb2.2}
\begin{tabular}{cll}
  \hline
  Parameters && Units\\
  \hline
  $\alpha$ & adult-tree mortality rate & $0.01\hspace{0.1cm}year^{-1}$\\
  $b$& seed dispersal rate & 0.05 $year^{-1}$\\
  $\delta$& tree competition coefficient & vary\\
  $f$ &lightning parameter & 0.33 $year^{-1}$\\
 $m$ & juvenile tree maturation rate & $0.1\hspace{0.1cm}year^{-1}$\\
 $r$ & grass recovery rate & $4\hspace{0.1cm}year^{-1}$\\
 $I$ & fire immunity & 0.3\\
 \end{tabular}
\end{table}


The core parameters used (see Table \ref{parnamesComb2.2}) are
based on those of \cite{Calabrese2010}, but with a few
modifications. The adult-tree death rate, $\alpha$, and the
lightning parameter, $f$, were changed according to
\citet[pag.\hspace{0.15cm}852]{Hanan2008} and
\citet[pag.\hspace{0.15cm}557]{Gignoux1997}, respectively. In
mesic savannas fire frequency is about once per year to once
every three years (\citet[pag.\hspace{0.15cm}852]{Hanan2008}
and \citet[pag.\hspace{0.15cm}557]{Gignoux1997}) so we set
$f=0.33$. The juvenile tree growth rate, $m$, was determined
from \citet[pg.\hspace{0.15cm} 219]{Hochberg1994}, so that in
the absence of fire, a juvenile tree takes on average 5 years
to reach the adult state. These parameters will be used through
the paper unless explicitly stated.

The spatially-limited dispersal of seeds from an adult to
neighboring sites, one of two key facilitative processes in the
model, occurs at the spatial scale of the first and second
neighborhood, whereas the main negative interaction, adults
competing with and inhibiting the development of juvenile
trees, occurs only at the scale of the first (near)
neighborhood.

This is the opposite situation as the one believed to
occur in extremely arid ecosystems,
namely short distance facilitation by local improvement of
water infiltration, and long range competition among plants mediated by
long superficial roots. In this last case, vegetation patterns are
expected to display a rather regular tree or patch spacing
\citep{Rietkerk2004,Rietkerk2008}. Our situation is more
appropriate for mesic savannas, and would tend to promote tree
clustering. But the occurrence of fire may alter the nature of
the interactions in a variety of ways, which we investigate in
the following.

\section{The tree-grass balance and tree extinction}
\label{sec:treegrass}

\begin{figure}[H]
\begin{center}
  $\begin{array}{cc}
      \includegraphics[width=5.9cm]{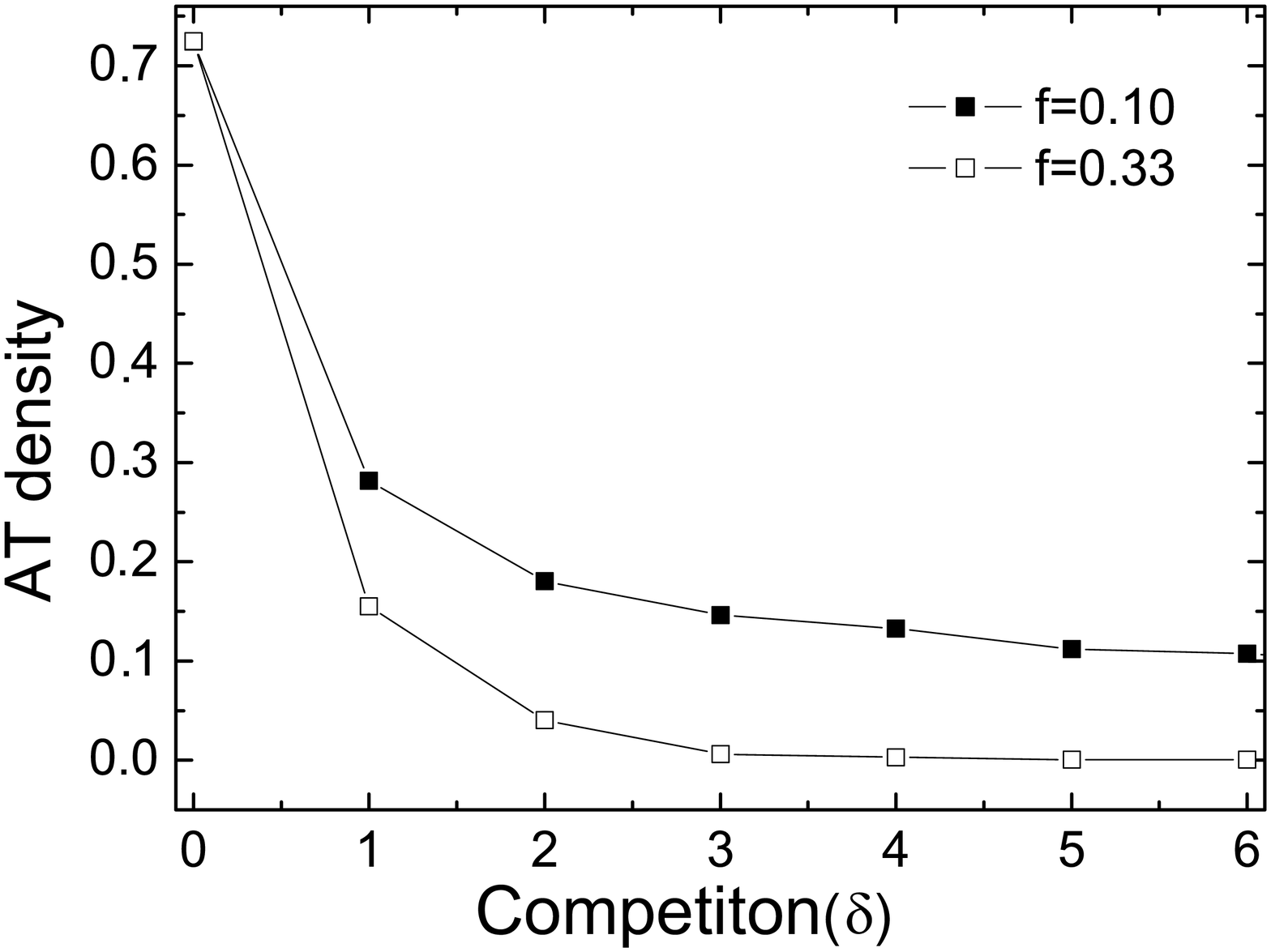} &\hspace{-0.1cm}
      \includegraphics[width=5.9cm]{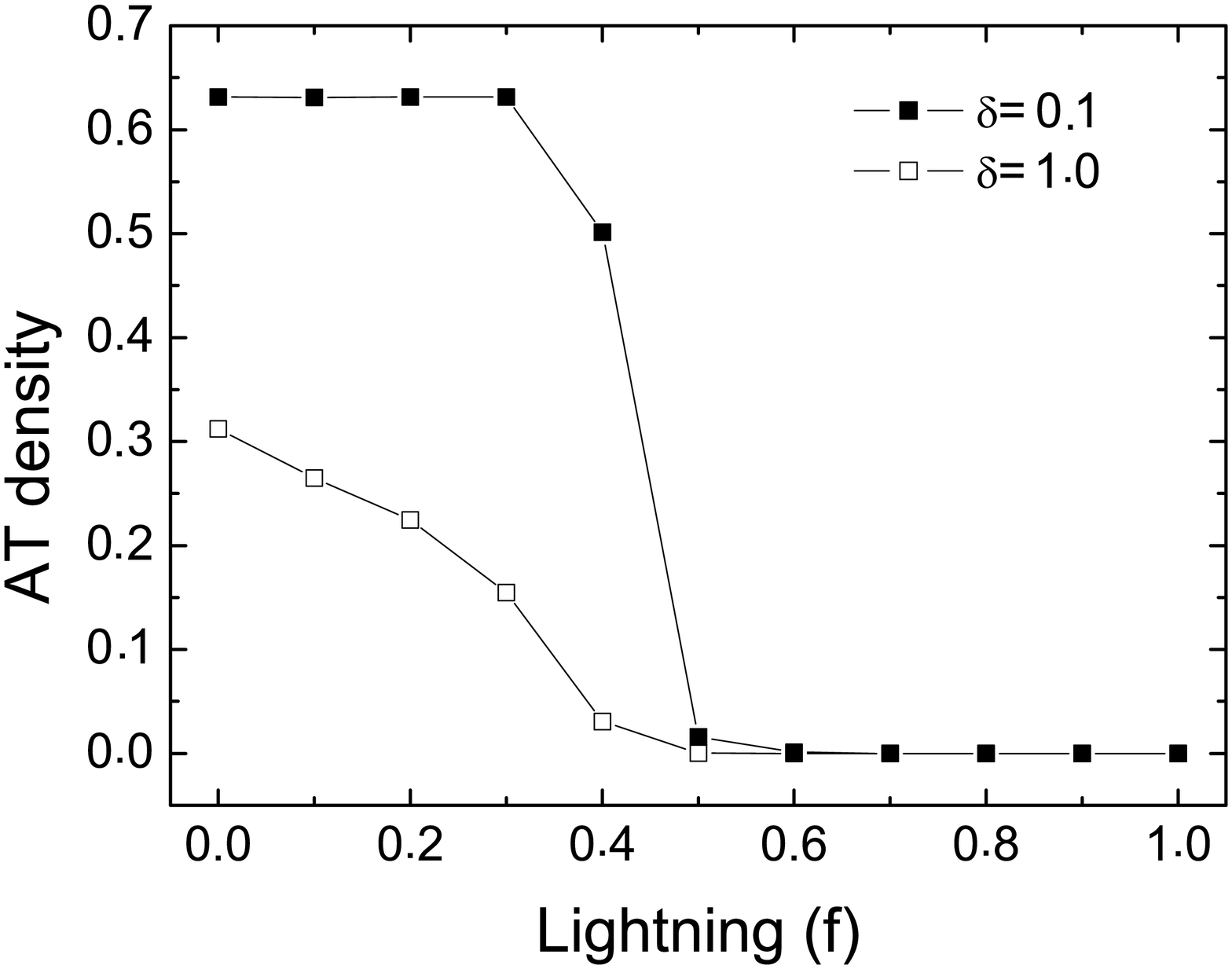}
      \\
   \end{array}
  $ \vspace{-0.5cm}
   \caption{Density of adult trees (i.e., the number of adult trees divided by the
   total number of lattice sites) versus competition ($\delta$, left graph) and versus
   lightning ($f$, right graph). Average over 500 snapshots in the long-time
   asymptotic state. Parameters are as in Table \ref{parnamesComb2.2}. Transition from the
   coexistence state to grassland is driven by increasing $\delta$ and/or $f$. }
   \label{p-density}
\end{center}
\end{figure}

As expected, stronger tree-tree competition shifts the
tree-grass balance in favor of grasses (Figure \ref{p-density},
left). The model was run for $5000$ years to ensure an
asymptotic state was reached in which we performed the
measurements described in Figure \ref{p-density}. Simulations
were performed to determine under which conditions a transition
from savannas to grassland occurs.

Similarly to the results \citet{Calabrese2010} obtained for
their fire parameter $\sigma$, the lightning frequency $f$
turns out to be the parameter with the strongest influence on
the savanna-grassland transition in the SFM: The right part of
Figure \ref{p-density} shows a phase transition from savanna to
grassland driven by increasing $f$ (\emph{c.f.} Figure 2 in
\citet{Calabrese2010}).

Frequent fires prevent juvenile trees from recruiting into the
adult population, and if this inhibition is strong enough, it
would eventually result in tree extinction. This mechanism and
the subsequent tree extinction it causes was implicitly
contained in the definition of the fire parameter in the SF of
\citet{Calabrese2010}. Here, the mechanism appears as a
consequence of the explicit presence of fire.

\section{Positive and negative effects of surrounding adult trees on
juvenile's: Protection vs. competition}
\label{sec:positivenegative}

In addition to affecting the tree-grass ratio, fire also
introduces the positive effect of juvenile tree protection by
surrounding adult trees. To analyze this effect in detail we
ran simulations in which only the fire propagation process
(step 7 in the above algorithm) occurs. The lattice is
initialized fully with grass except for one unique site
occupied by a juvenile tree, and a number of adult trees, from
1 to 8, occupying random positions in the Moore neighborhood of
the JT. Given this initial condition, one sparking is allowed
so that a lattice site chosen randomly among the G sites burns
and fire begins to propagate. Step 7 in the SFM algorithm is
repeated until fire disappears.

\begin{figure}[H]
\begin{center}
  $\begin{array}{cc}
    \includegraphics[width=5cm]{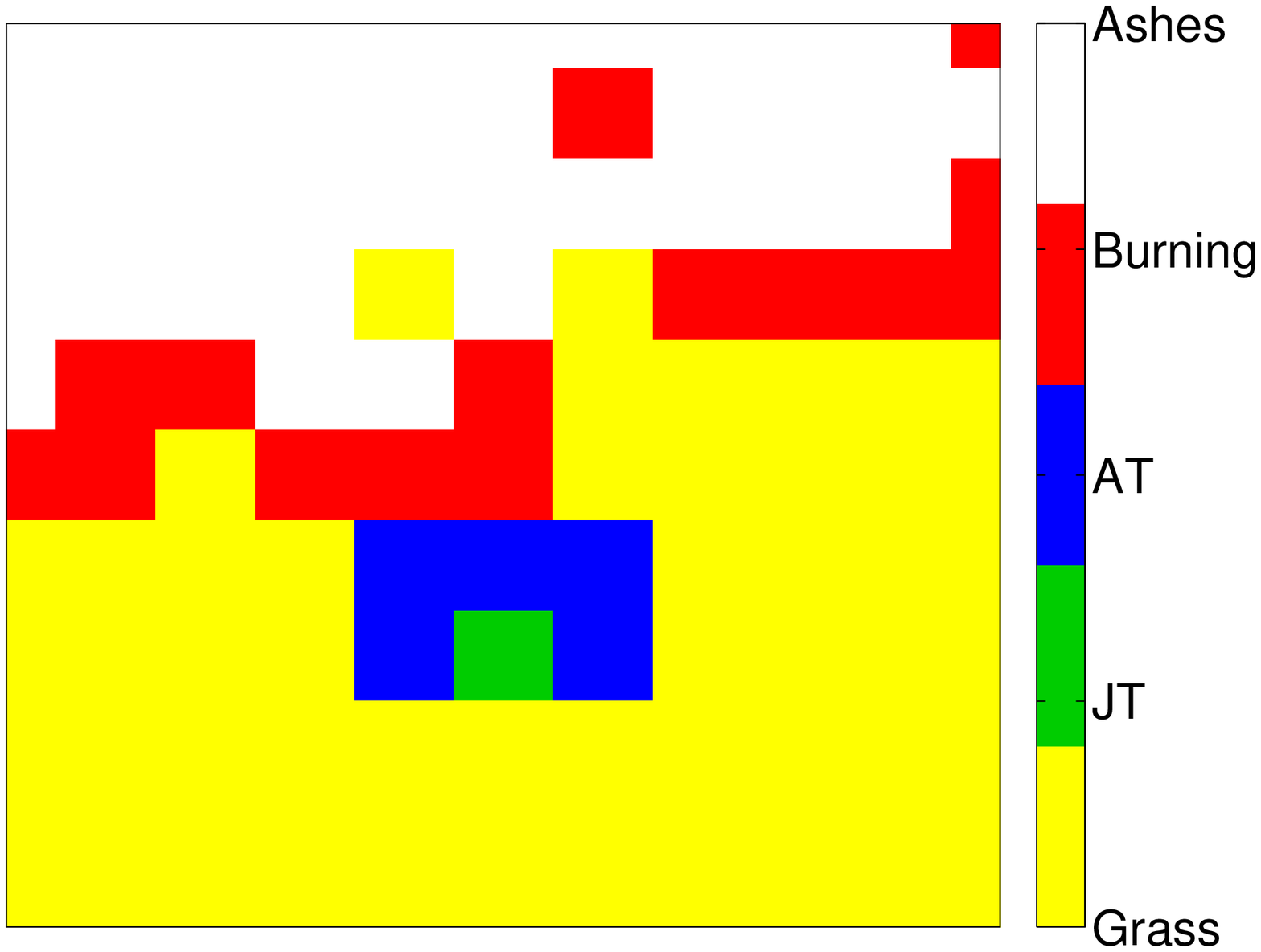}
    & \includegraphics[width=5cm]{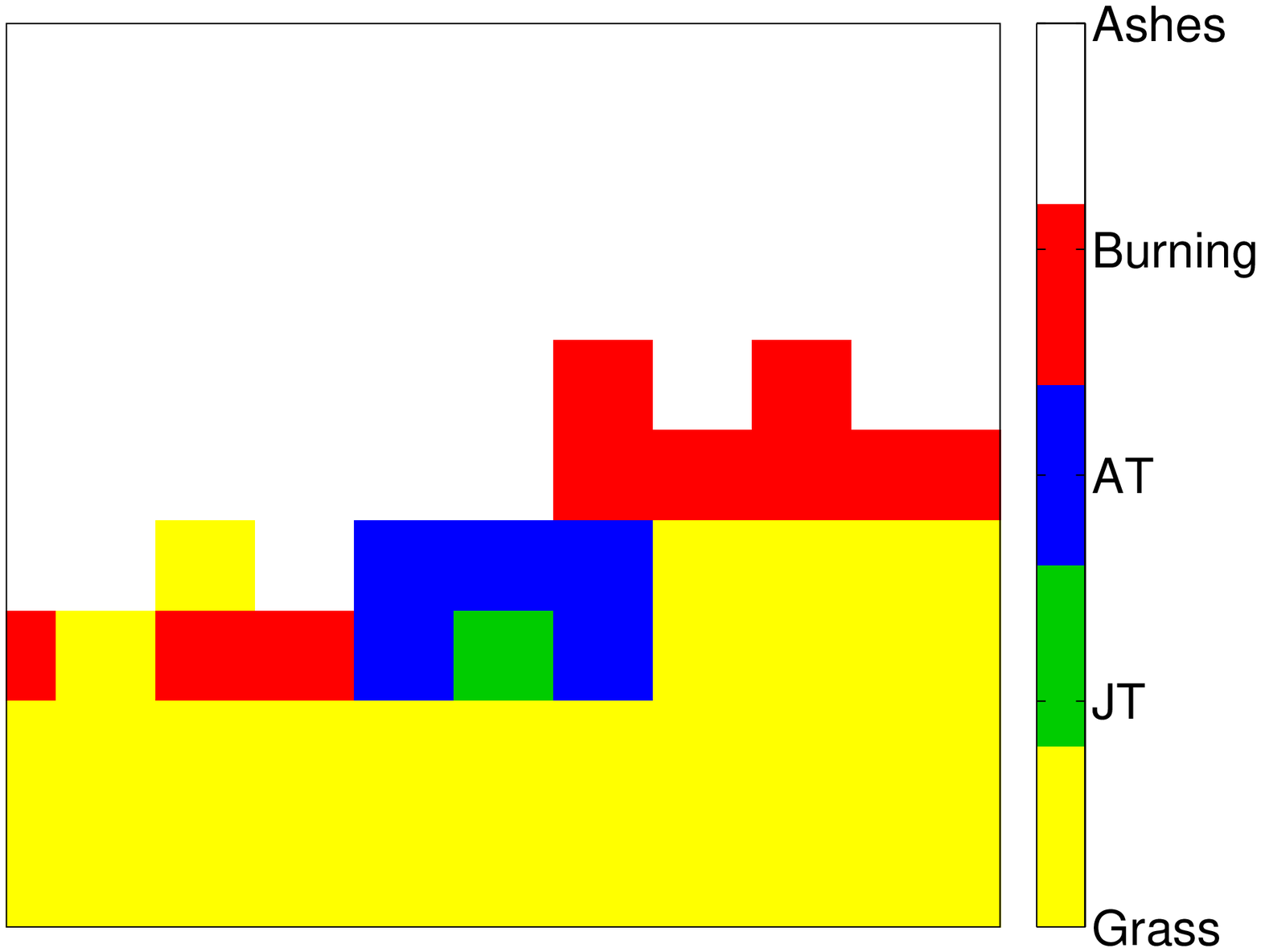} \\
    \includegraphics[width=5cm]{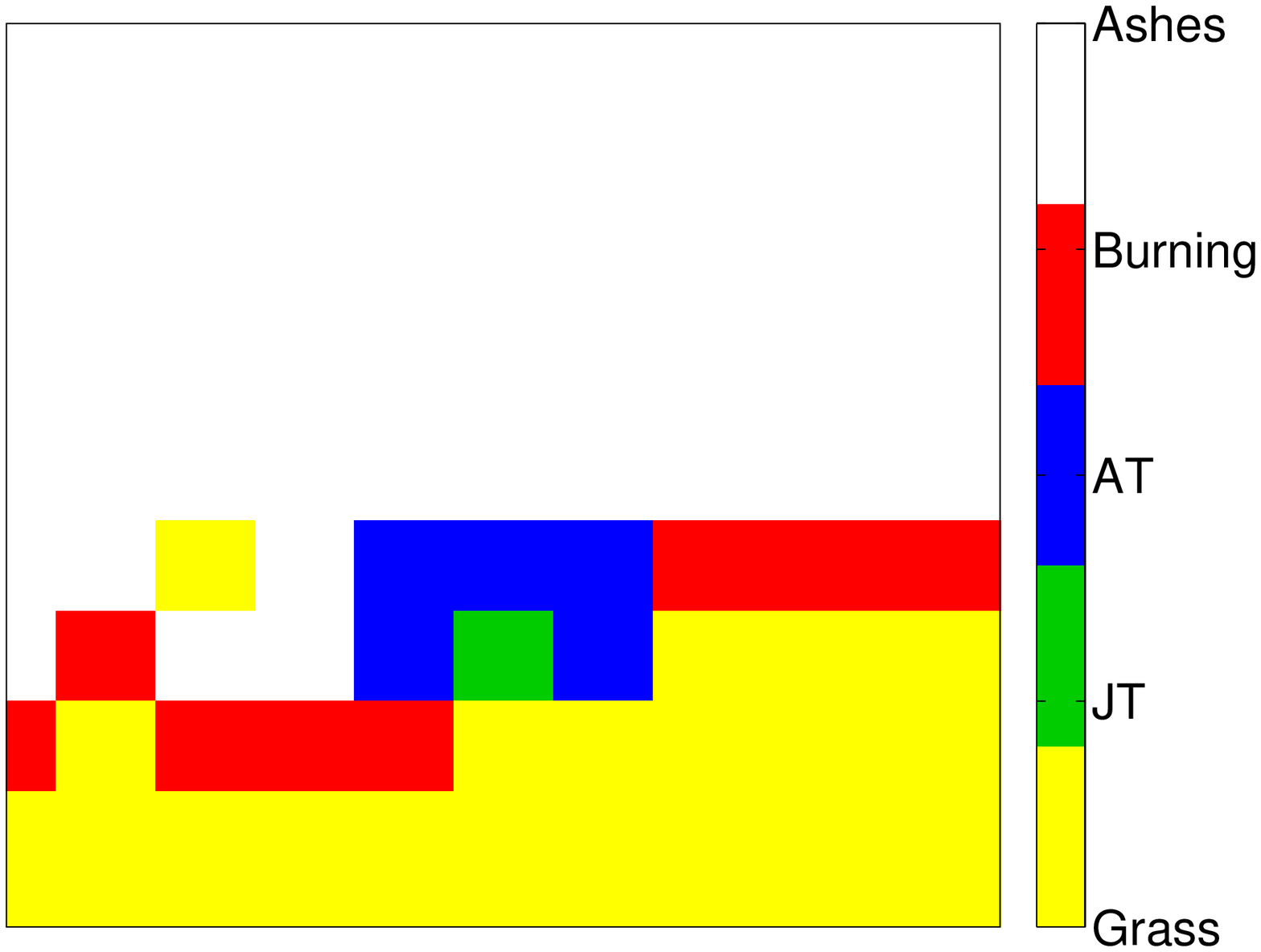} &
    \includegraphics[width=5cm]{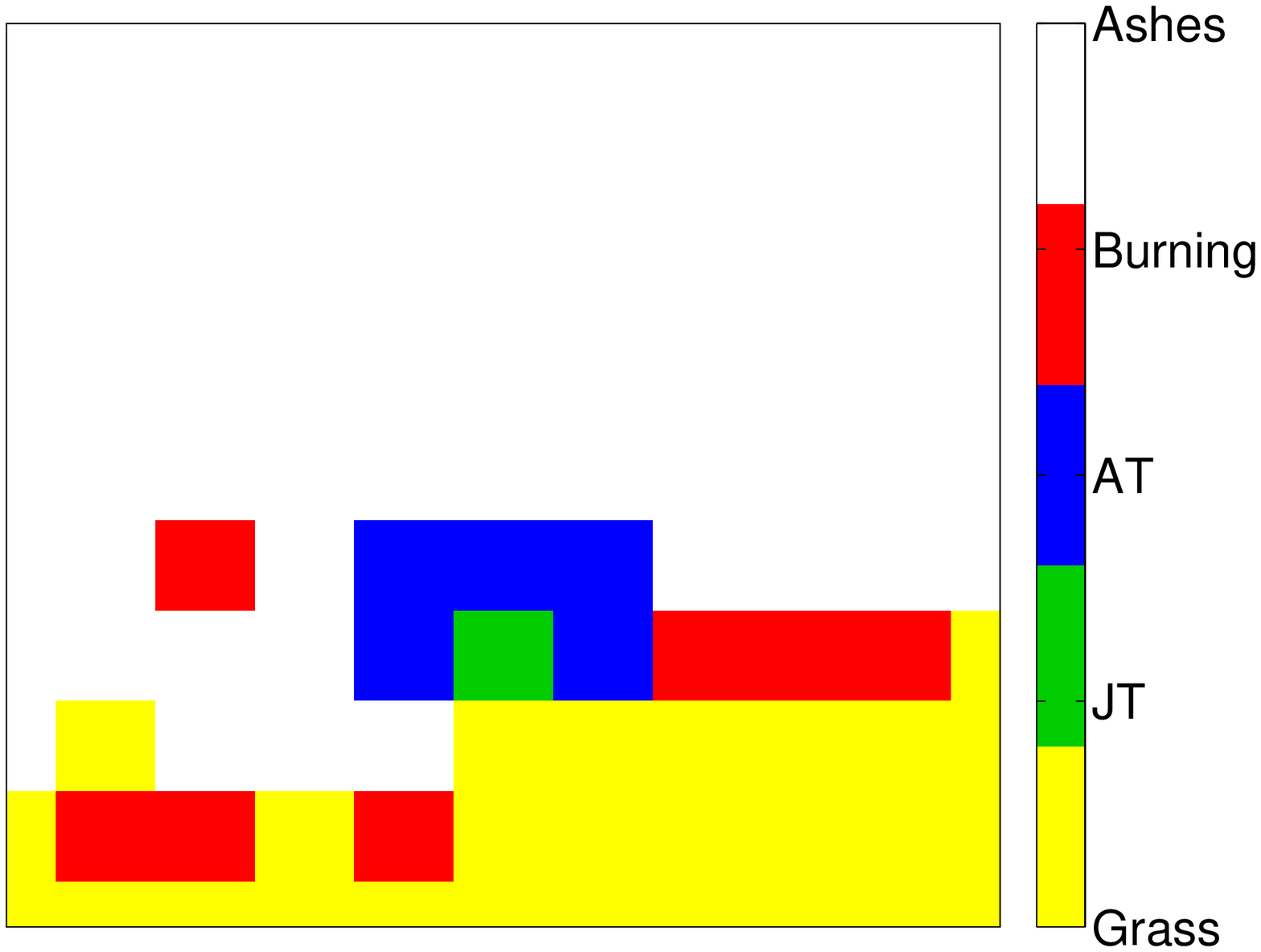} \\
    \includegraphics[width=5cm]{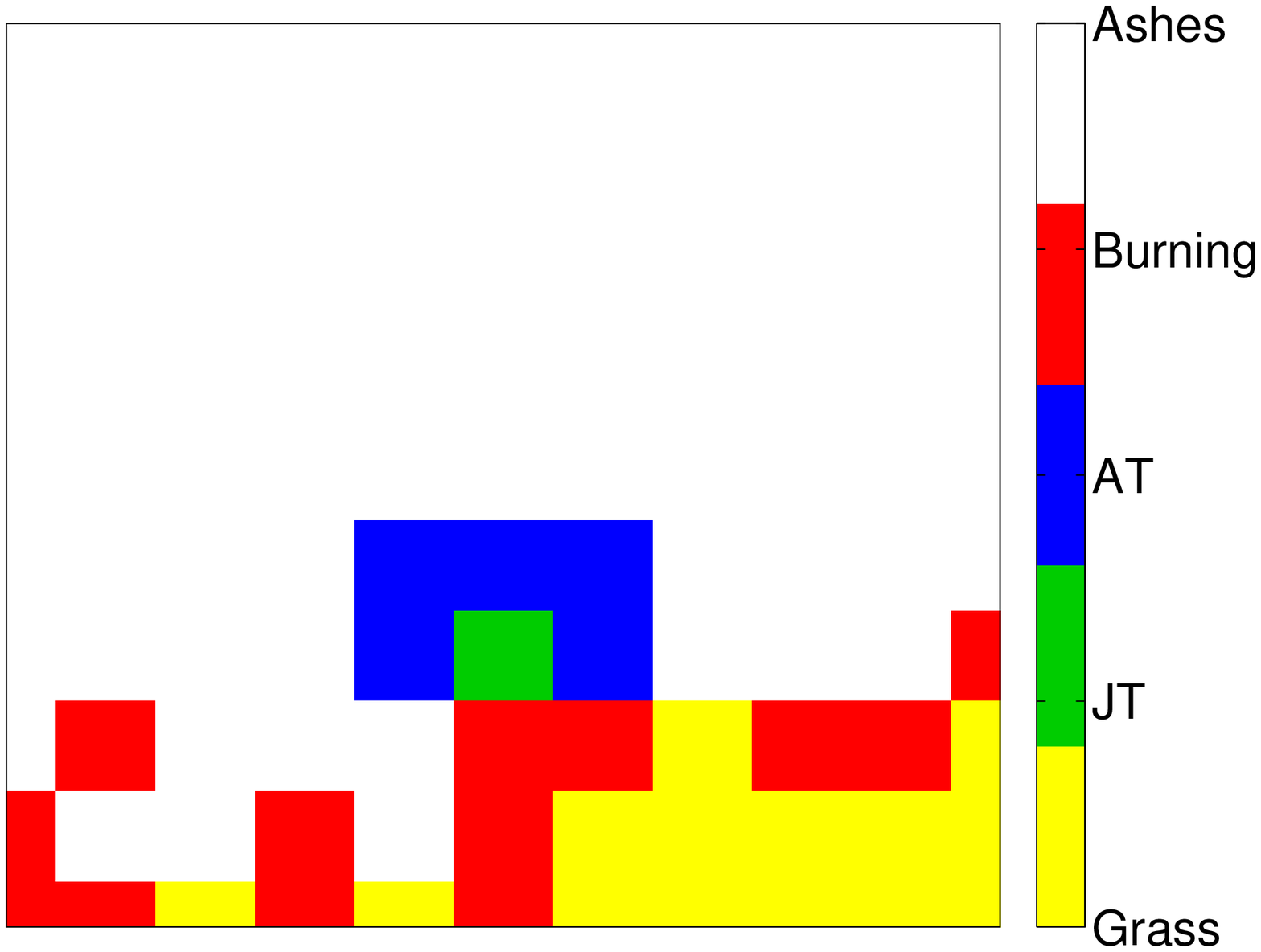}&
    \includegraphics[width=5cm]{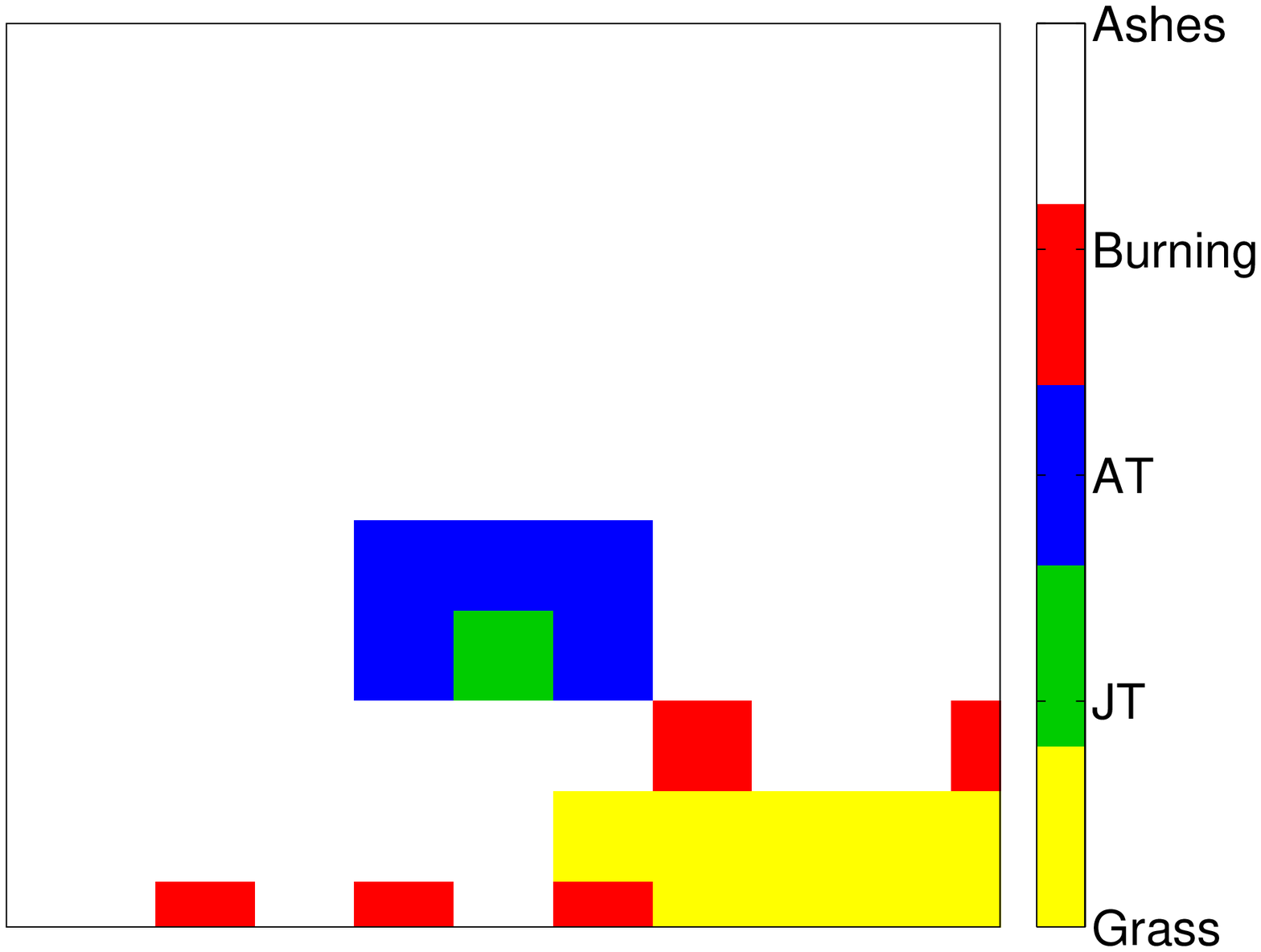}
  \end{array}$
   \caption{Protection effect: Selected snapshots from an example simulation with 5 adult
   trees (blue) surrounding a juvenile (green). Immunity parameter $I=0.3$.
   Time runs from left to right and then from the upper to the
   lower row. A fire front (red) advances downwards, converting grass
   (yellow) into ashes (white), but the juvenile survives. Only a $10\times10$
   area of the whole $200\times200$ lattice is shown.}\label{protection-snapshots}
\end{center}
\end{figure}

Juvenile trees sufficiently protected by adult trees will not
burn. An example is seen in Fig. \ref{protection-snapshots},
where the pass of a fire front does not affect a juvenile
protected by five adult trees.

\begin{figure}[ht]
\begin{center}
  \includegraphics[width=7cm]{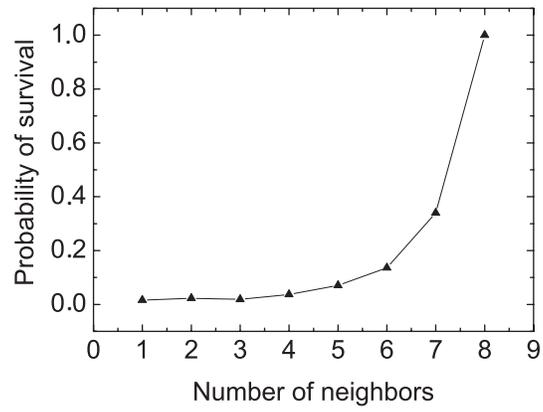}\\
  \caption{The protection effect: The probability $P_F^{Surv}(S_1)$ of a juvenile surviving
  one fire as a function of the number of surrounding adult trees in its first
  neighborhood. This probability has been obtained from 1000
  realizations of the process in which fire is initiated at one grass site, as described in
  the text, using immunity $I=0.3$.}\label{protection}
\end{center}
\end{figure}

We quantified this effect by repeating the burning
protocol 1000 times for each number of AT neighbors, from 1 to
8. The sparking site and the position of the surrounding
neighbors is changed randomly in each of these realizations. The resulting
survival probability $P_F^{Surv}(S_1)$ is shown in Fig.
\ref{protection}. The protection provided by an increasing number
of ATs is clearly seen when the immunity parameter is not too
small. For very small $I$, protection is only effective when
the juvenile is completely surrounded by adults, i.e. $S_1=8$.

To better quantify the impact of the protection effect on
juvenile survival and recruitment, we now estimate how the
number of adult trees $S_1$ in the near neighborhood of a site
affects the recruitment probability $P_R(S_1)$ defined as the
probability that a grass site becomes successfully colonized by
a tree seed during a given time-step and the resulting JT
survives successive fires and becomes an adult. This
probability is a product of several factors.

First, the grass site should receive in that time step a seed
from the adult trees in the near or in the far neighborhood
(the numbers of adult trees there are $S_1\in(0,8)$ and
$S_2\in(0,16)$, respectively). This is given by
$P_s(S_1,S_2)=1-(1-b \Delta t/24)^{S_1+S_2}$. Then, the seed
establishes as a juvenile tree and must survive competition
during successive time steps, which is given by the factor
$P_C^{Surv}=\exp{(-\delta S_1)}$. Since $m^{-1} $ is the
average growth time from juvenile to adult tree, $(m\Delta
t)^{-1}$ time steps occur during growth, and
$\exp{(-\frac{\delta S_1}{m \Delta t})}$ is the total survival
factor to adulthood under competition.

Finally, the growing JT should resist the first and successive
fires occurring during its growing time $m^{-1}$. The
probability of surviving a single fire is the function
$P_F^{Surv}(S_1)$ numerically calculated and shown in Fig.
\ref{protection} for the case in which the focal site is
surrounded by $S_1$ trees in the near neighborhood (i.e.,
$S_2=0$). An estimation of the probability surviving successive
fires, which neglects any correlations arising from successive
fire fronts and from the lattice configuration beyond the
immediate neighborhoods, would be
$\left(P_F^{Surv}(S_1)\right)^{f/m}$, where $f/m$ is the
expected number of fires suffered by the JT during its growing
time $m^{-1}$. The probability $P_s(S_1,S_2)$ depends on the
number of AT both in the near and in the far neighborhood. For
consistency with the calculation of $P_F^{Surv}(S_1)$ we will
take $S_2=0$. This (as in the case of $P_F^{Surv}(S_1)$) will
underestimate the probability of establishment, survival and
recruitment, as the trees in the far neighborhood do not
compete with the central one. In this way we will obtain an
estimation of the recruitment probability $P_R(S_1)$ that is
smaller than the exact one. Thus, if this function shows
positive effects of surrounding adult trees, the exact result
must be larger, since our approximation is obtained in a {\sl
worst case} situation.

Summarizing all the factors above, with $S_2=0$, our estimation
of the recruitment probability of a grass site surrounded by
$S_1$ adult trees is

\begin{equation}
P_R(S_1) \approx \left[1-\left(1-\frac{b \Delta t}{24}\right)^{S_1}\right]
e^{-\frac{\delta S_1}{m \Delta t}}
\left(P_F^{Surv}(S_1)\right)^{\frac{f}{m}}
\label{recruiting}
\end{equation}

\begin{figure}
  \centering
      \includegraphics[width=7cm]{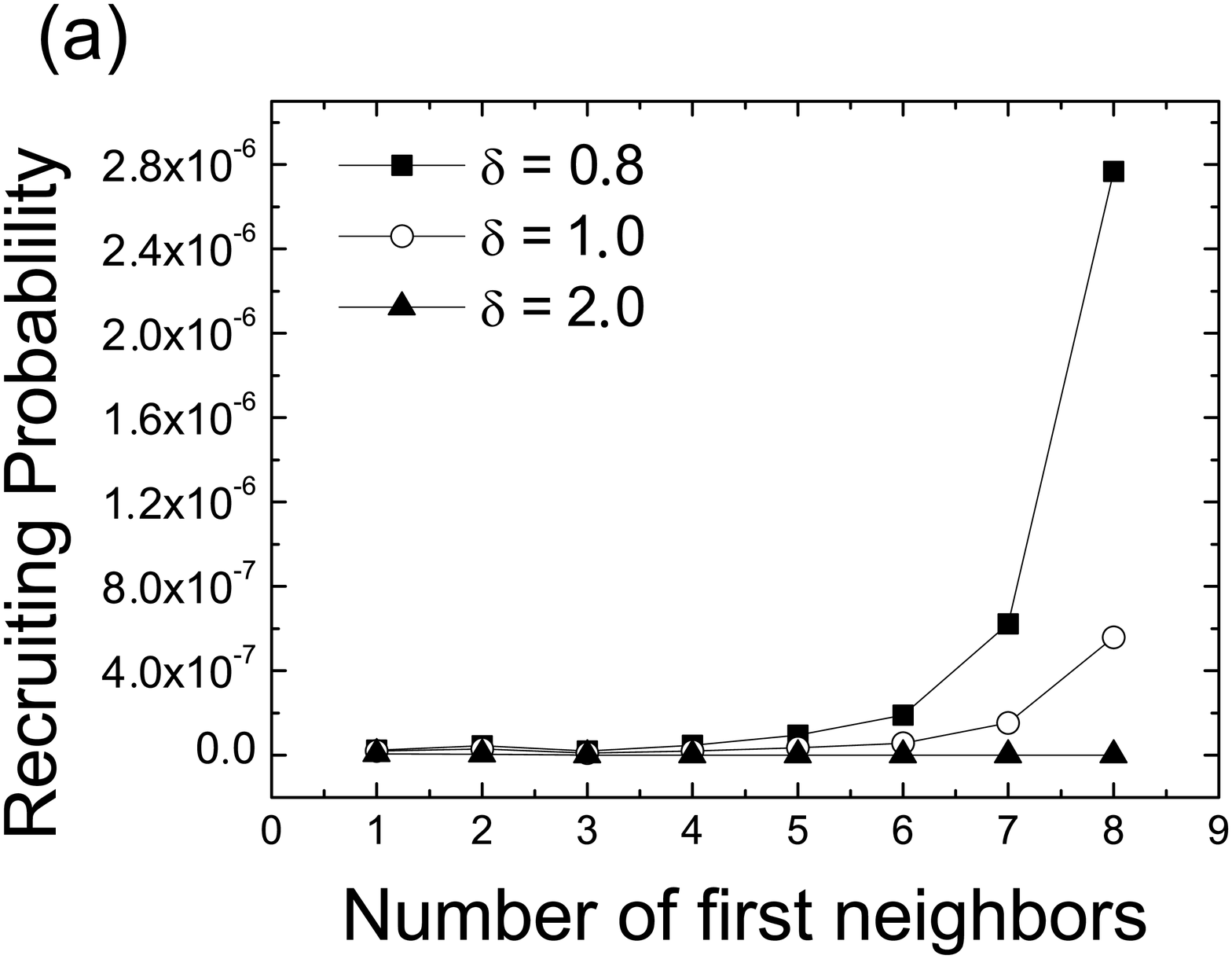} \\
     \includegraphics[width=7cm]{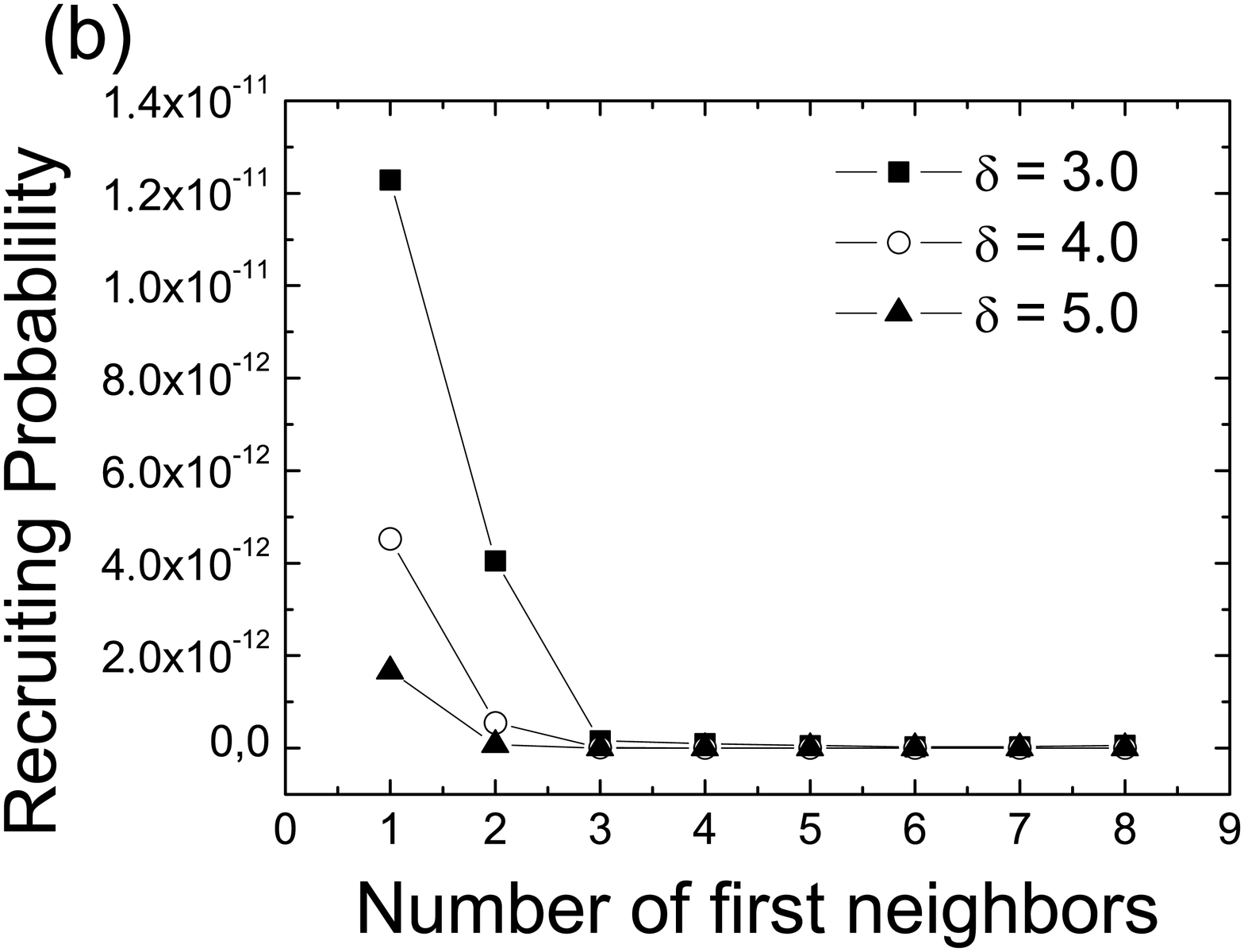} \\
     \includegraphics[width=7cm]{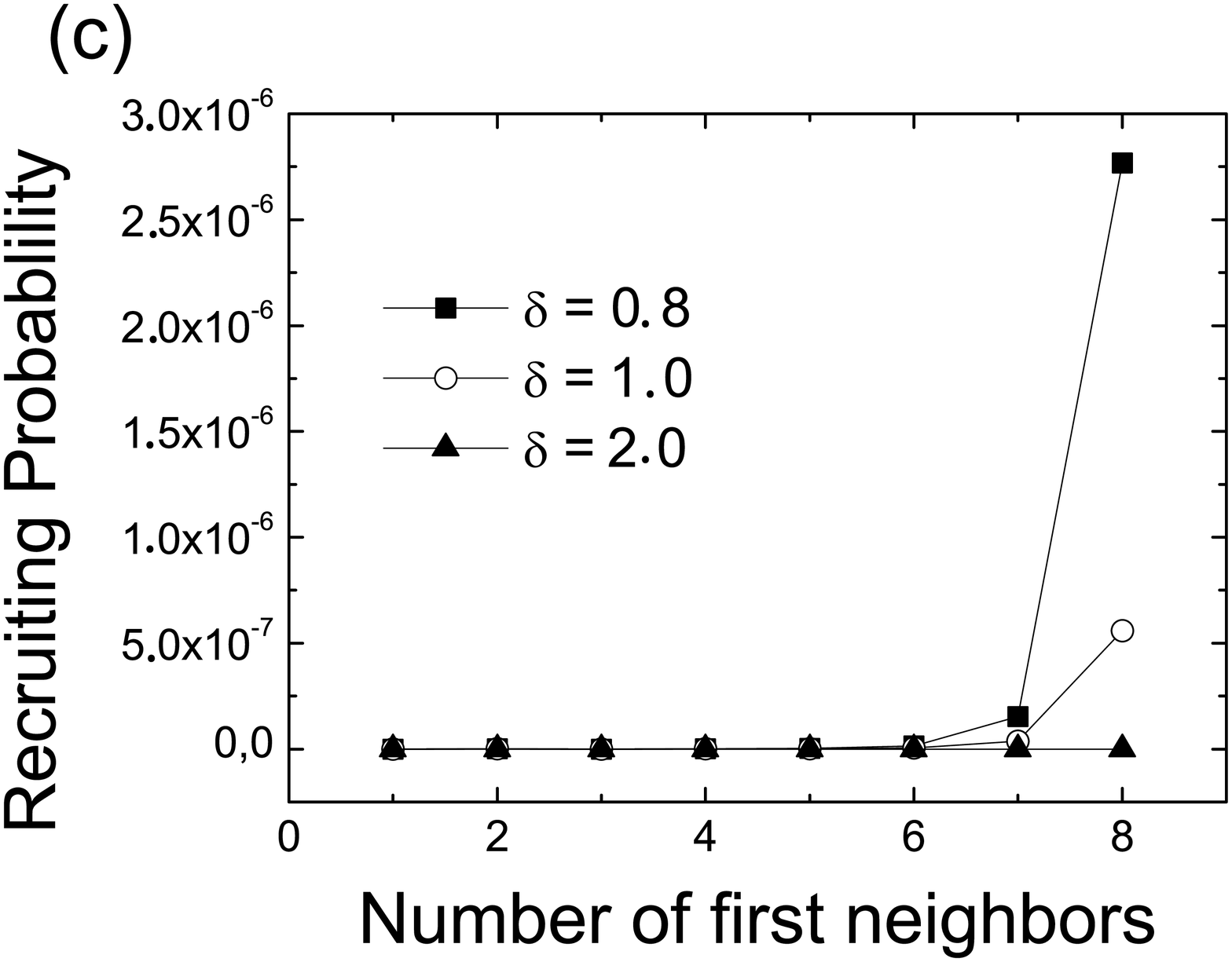}\\
  \caption{Estimation of the recruiting probability $P_R(S_1)$, as a function
  of the number of adult trees $S_1$ in the near neighborhood,
  from Eq. (\ref{recruiting}), showing the positive and negative effects of these neighbors.
  (a) and (b) $I=0.3$, $f=0.33\ year^{-1}$ (triennial fire). (c) $I=0.3$, $f=0.2\ year^{-1}$
  (pentannual fire).
  }
  \label{botheffects}
\end{figure}

This is plotted in Fig. \ref{botheffects}, and reveals both the
positive and the negative effects of the presence of
neighboring trees (but remember that the positive effects are
underestimated). For medium values of the competition parameter
and above four neighboring adult trees the positive protective
effect of fire (in combination with local dispersal) overcomes
the negative effect of direct competition (see Figure
\ref{botheffects}(a) and (c)), but for high values of
competition the negative effect predominates (see Figure
\ref{botheffects}(b)). For frequent fire, however, the
protection effect is no longer effective.

\section{Clustering patterns}
\label{sec:clusters}

\subsection{Tree spatial pattern under different fire scenarios}

We characterize spatial patterns of adult trees by the pair
correlation function \citep{Dieckmann2001}, $g(l)$:
\begin{equation}\label{g}
    g(l)=\frac{\rho_{AA}(l)}{(\rho_A)^2}
\end{equation}
\noindent where $\rho_{AA}$ is the proportion of pairs of adult
trees at a distance $l$ (with respect to the total number of
pairs of sites at that distance) and the denominator is the
expected value of this proportion under a random distribution
with the density of the adult trees $\rho_A$. At large
distances $g(l)$ is expected to approach 1, as correlations
indicating a departure from random distribution would decay.
For short distances, $g(l)$ characterizes how the trees are
packed together \citep[see][chap. 14]{Dieckmann2001}, values
higher than 1 indicating a proportion of pairs at that distance
greater than in the random case (clustering), and a smaller
proportion indicated by values of $g$ smaller than 1 (revealing
a more regular spacing). We will not use the Euclidean distance
for $l$ but instead we will measure $l$ in number of cell
layers so that $g(1)$ and $g(2)$ will denote the pair
correlation function for the first and for the second Moore
neighborhood, respectively.

\begin{figure}[ht]
  $\begin{array}{@{}c@{}@{}c@{}@{}c@{}}
    \includegraphics[width=4cm]{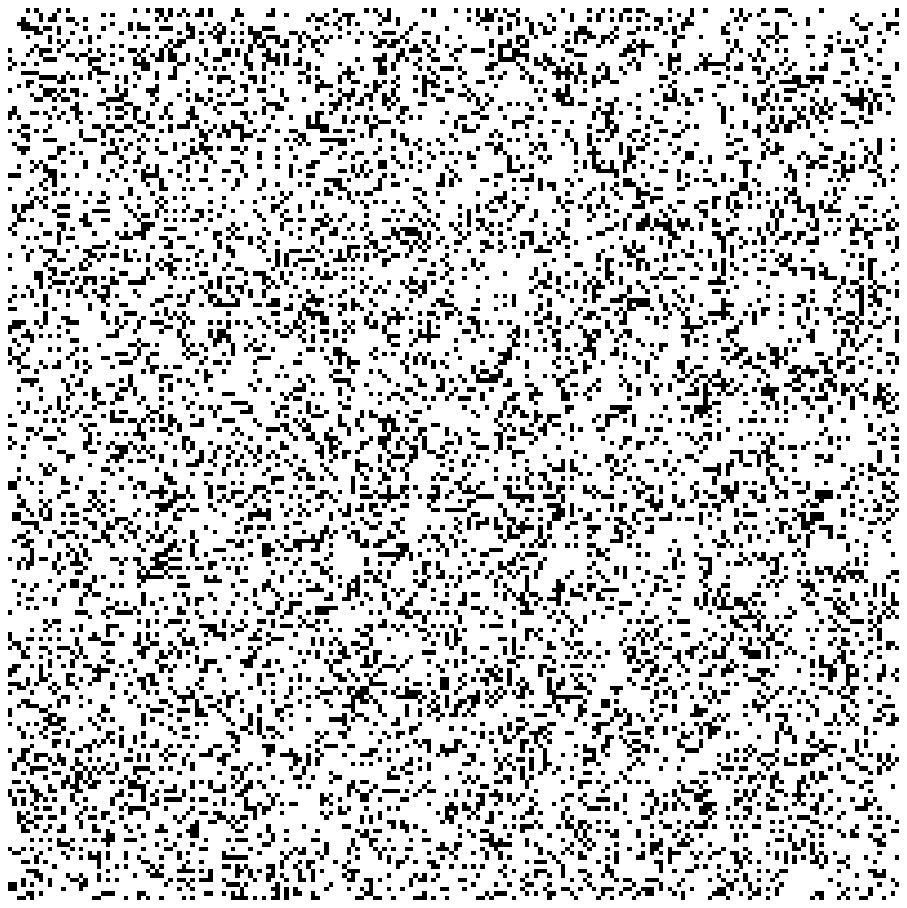} &
    \includegraphics[width=4cm]{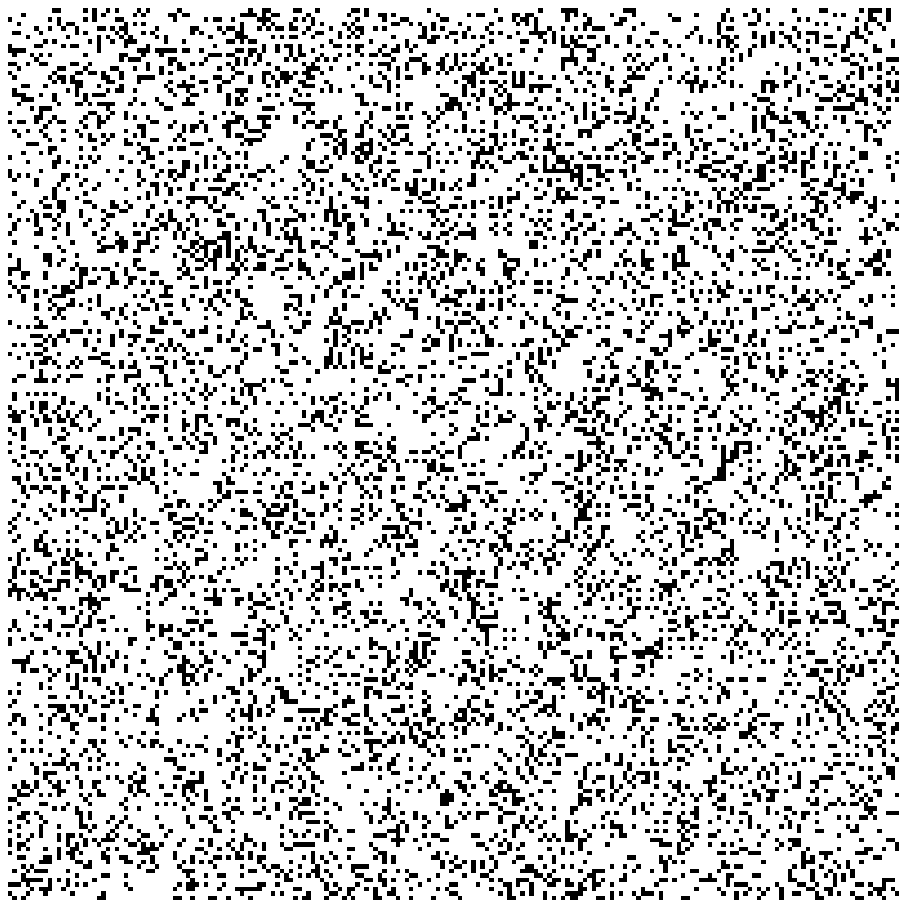}  &
    \includegraphics[width=4cm]{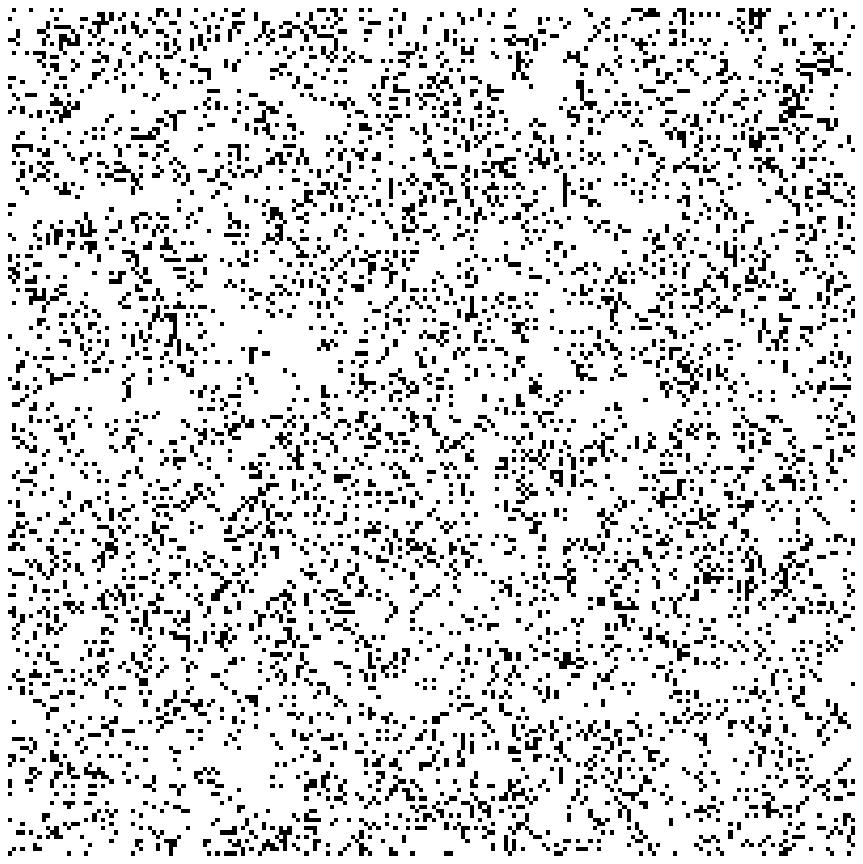} \\
      \multicolumn{1}{c}{\makebox[0.5in][r]{\bf Regular}}
      &\multicolumn{1}{c}{\makebox[0.45in][r]{\bf }} &
     \multicolumn{1}{c}{\makebox[0.45in][r]{\bf Clumped}}\\
     \includegraphics[width=5cm]{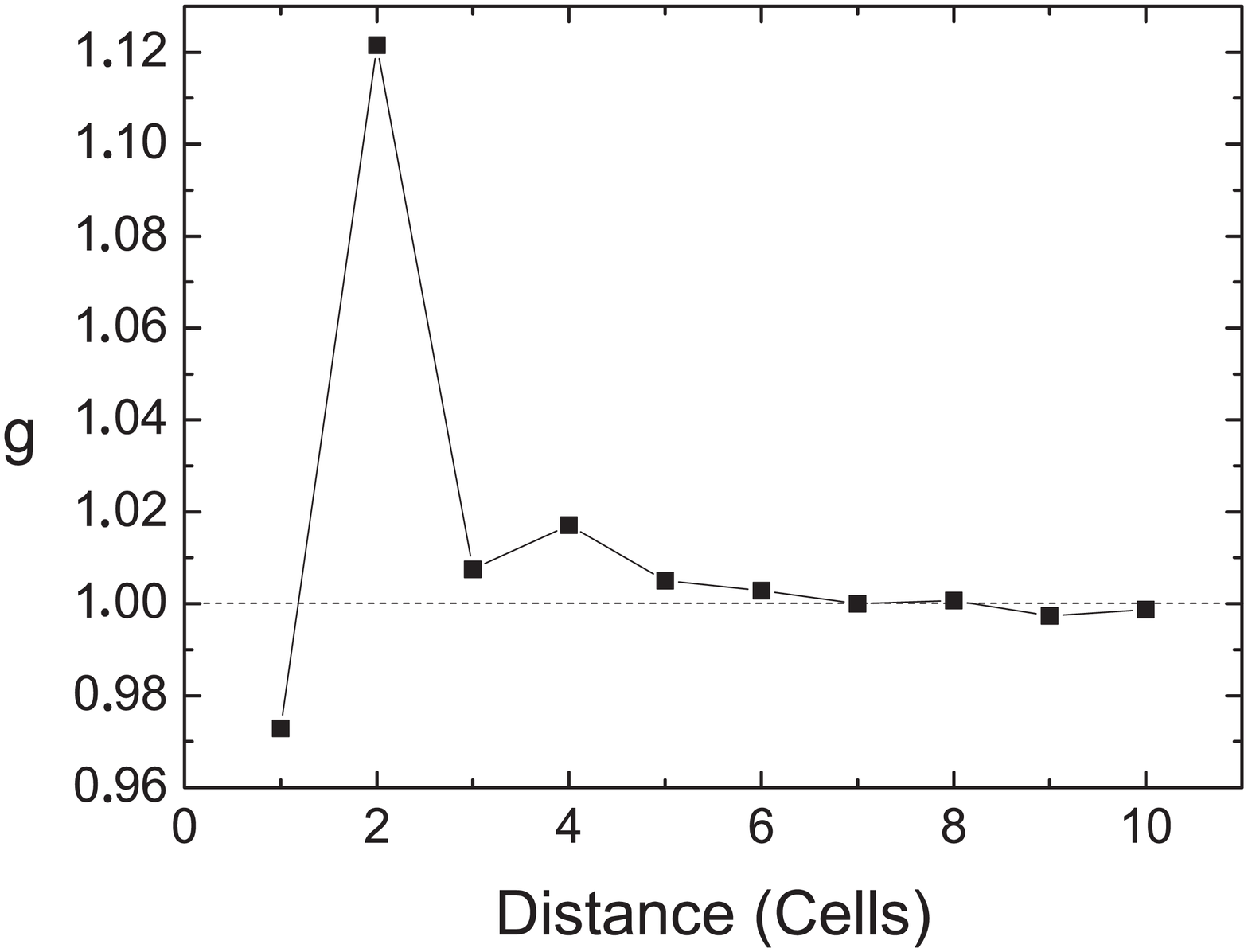} &
     \includegraphics[width=5cm]{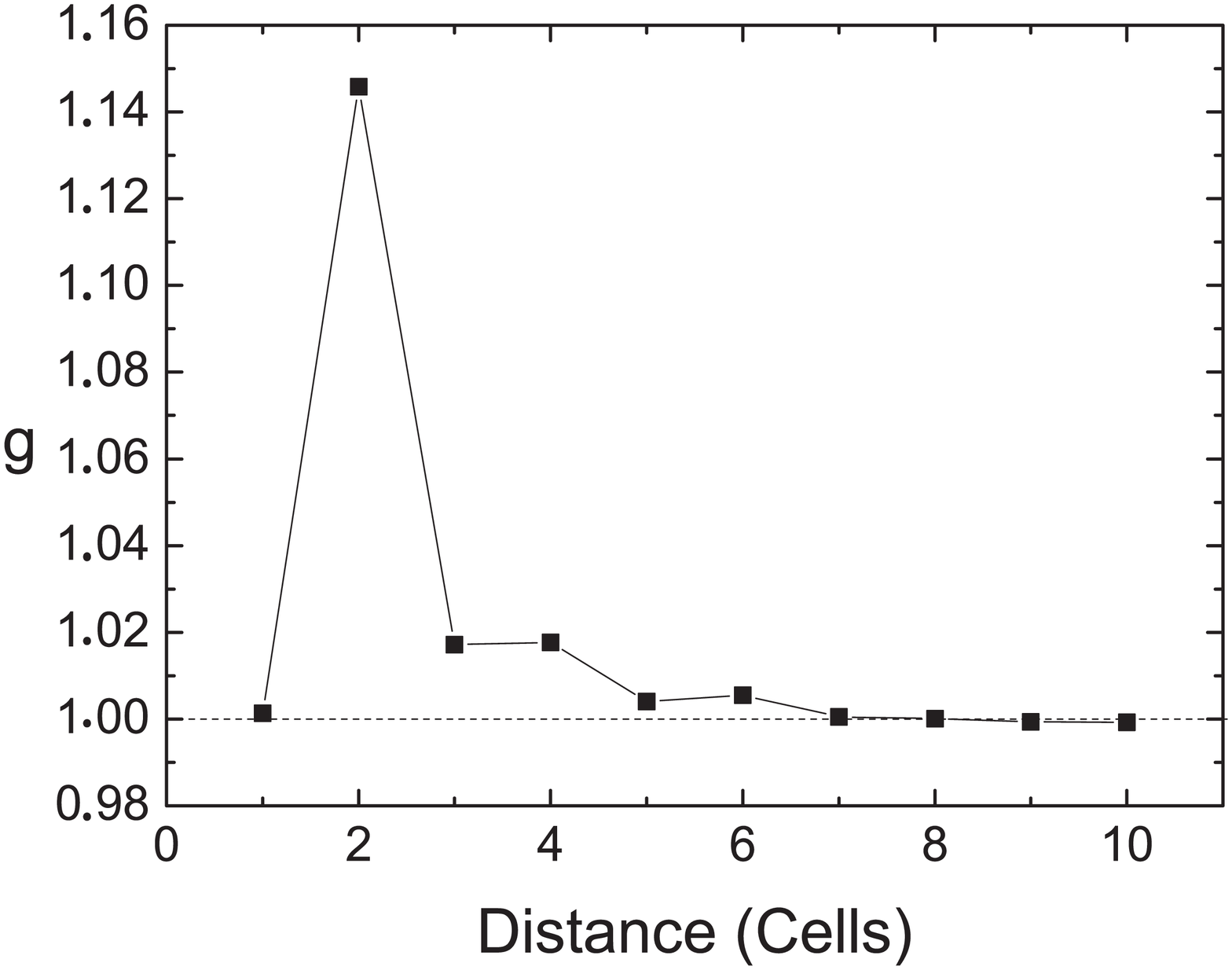} &
      \includegraphics[width=5cm]{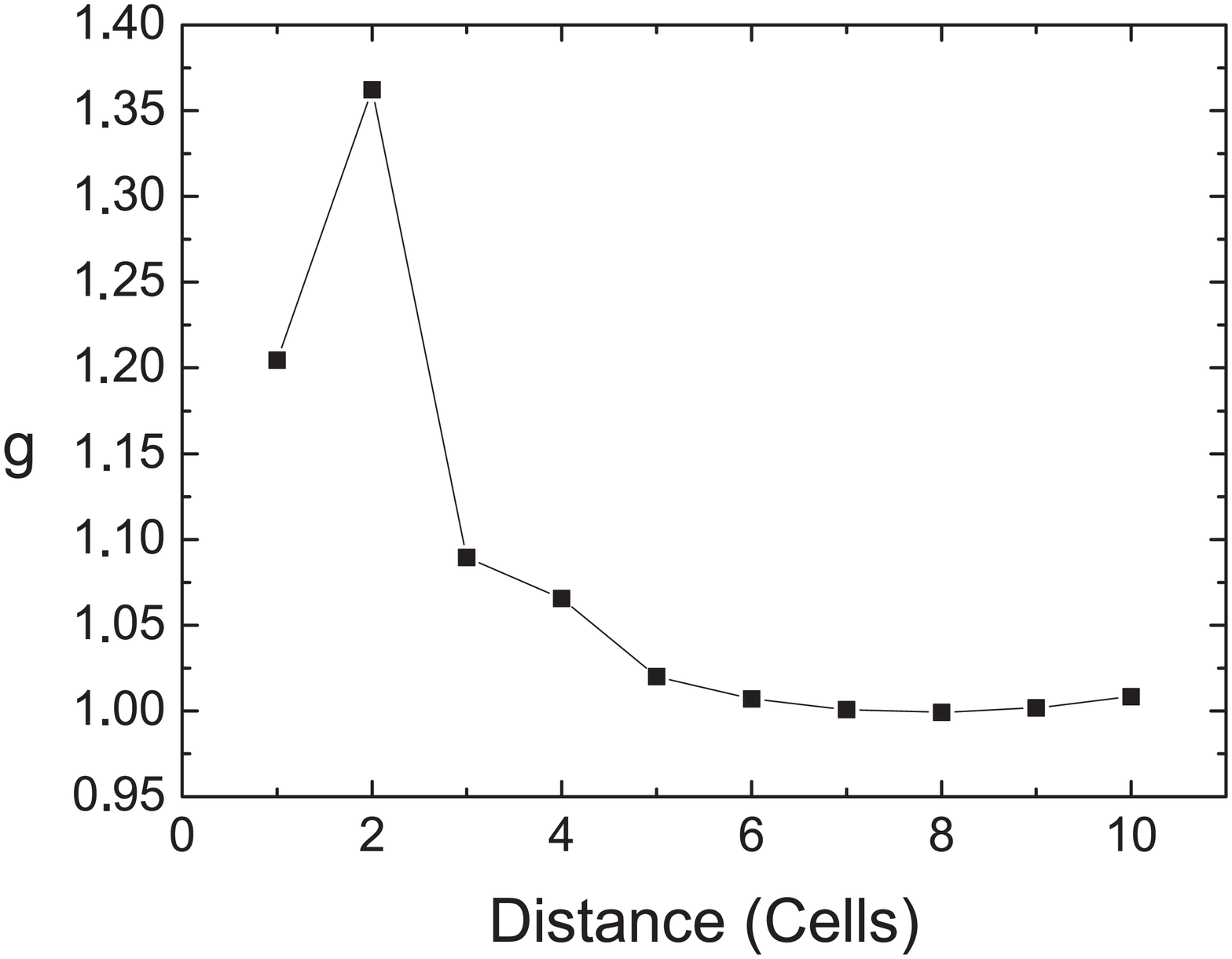}
  \end{array}$
  \caption{Patterns in the SFM. Parameters as in Table \ref{parnamesComb2.2}
  and $\delta=0.01$. Regular case: $f=0.19$. Clumped case:
  $f=0.31$.
  The central panel shows an intermediate
  state ($f=0.223$) in which $g(1)=1$, which indicates the same number of AT near
  pairs as in a random case.}\label{patterns}
\end{figure}

Comparison of Figure \ref{patterns} with the results of
\citet{Calabrese2010} shows that all the patterns found in the
SM are also present here. Some features of the patterns
can be understood from the fact that there is direct
competition only between nearest neighbors, whereas the
facilitation effect of local seed dispersal reaches first and
second neighbors. In consequence, all these patterns have an
enhanced probability of ATs having other ATs as second
neighbors (far Moore neighborhood), as seen by the high value
of $g(2)$. As in \cite{Calabrese2010}, two types of
configurations are distinguished by having a value of $g(1)$
smaller or larger than 1, i.e. smaller or larger proportion of
ATs in the near neighborhood than the one expected from a
random distribution. The balance between positive and negative
tree-tree interaction effects determines these values. The case
$g(1)<1$ is a {\sl regular} case in which trees appear more
regularly spaced than in the random case. The case $g(1)>1$ is
a {\sl clumped state}, in which, although the density of
near-neighbor pairs is still smaller than the one of
far-neighbor pairs, it is larger than in the random case. The
transition between the two states was governed by the parameter
$\sigma$ in the SM, which controls the probability of surviving
fire. Here, this transition is determined by the explicit fire
parameter $f$.

\begin{figure}[ht]
\centering
 $ \begin{array}{cc}
 \multicolumn{1}{c}{\makebox[-1.5 in][r]{\bf (a)}}
 &\multicolumn{1}{c}{\makebox[-1in][r]{\bf (b)}}\\ [-0.3cm]
  \includegraphics[width=3cm]{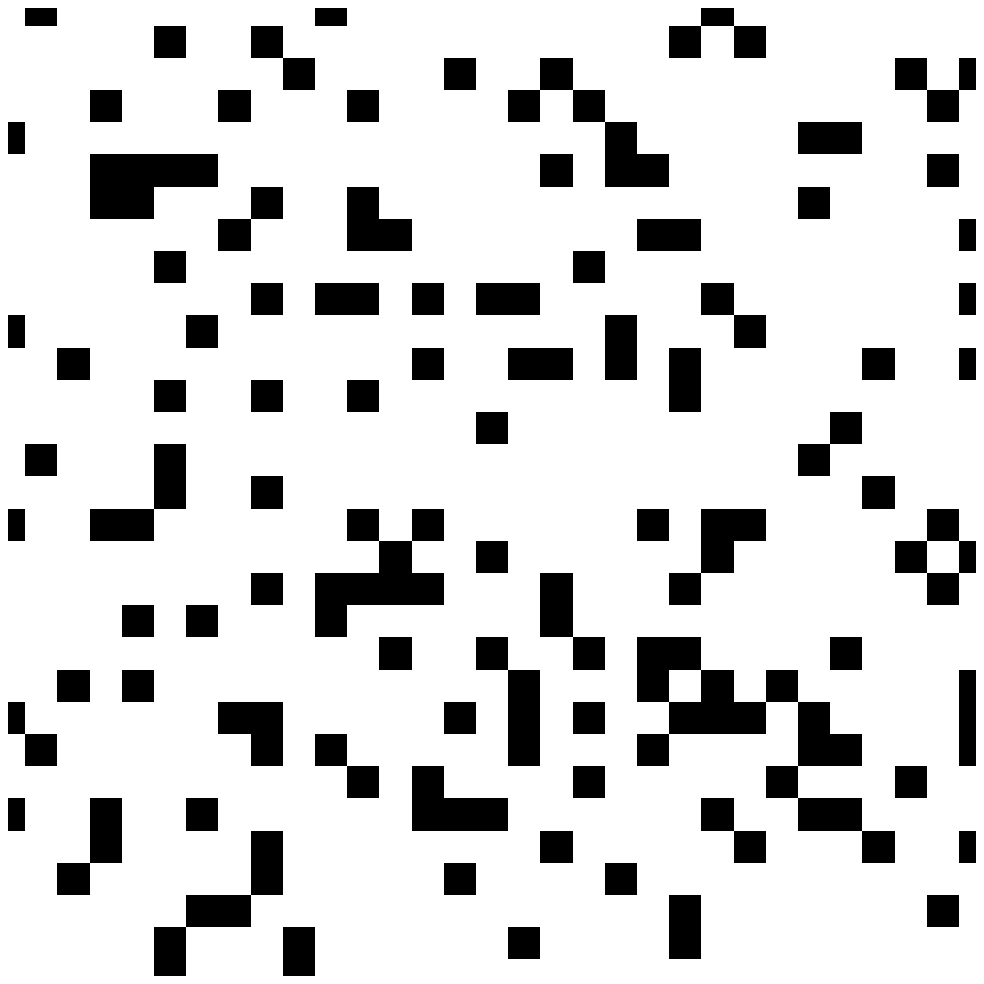} & \hspace{1.8cm}
  \includegraphics[width=4.5cm]{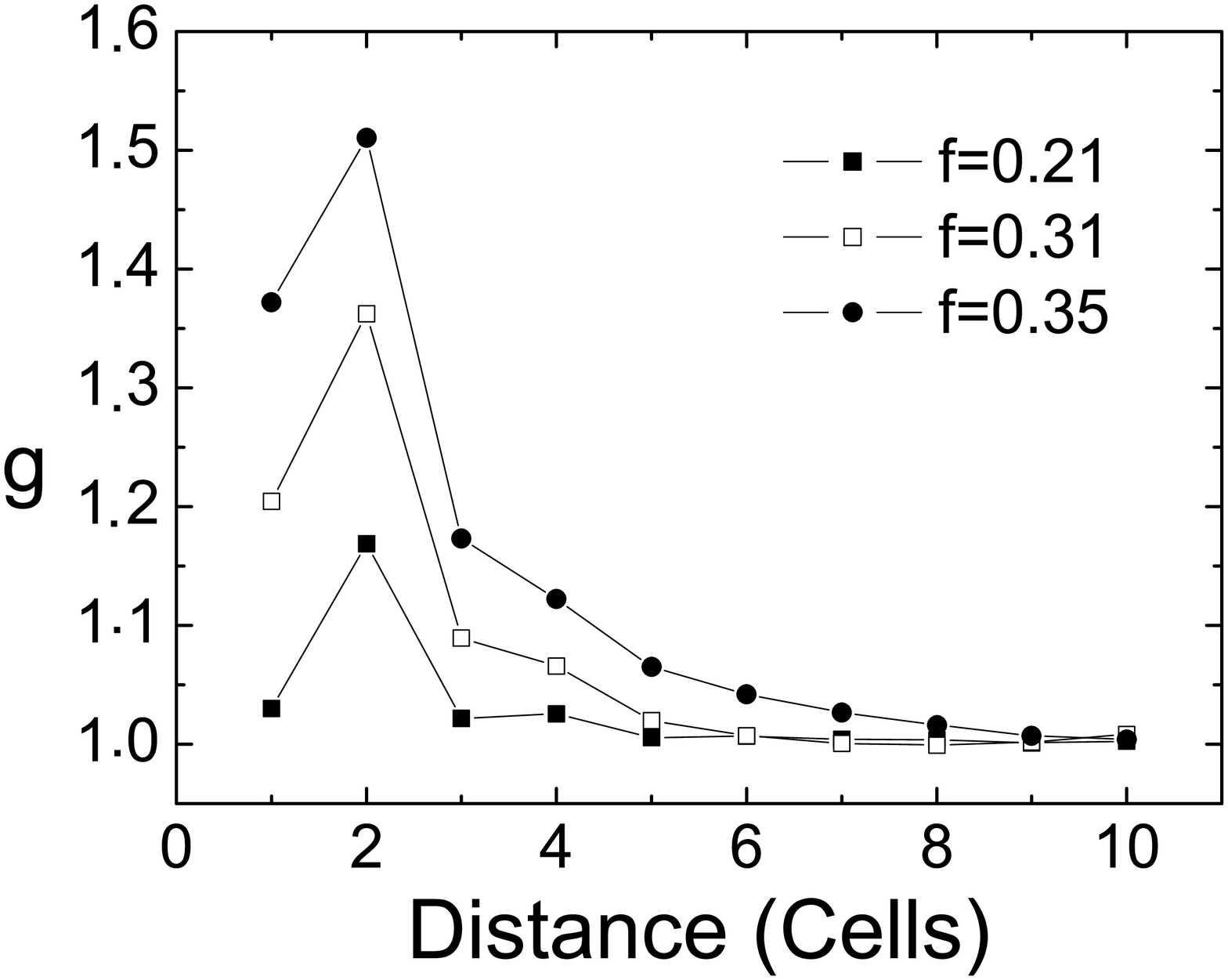}\\[0.5cm]
   \multicolumn{1}{c}{\makebox[-1.5 in][r]{\bf (c)}}
   &\multicolumn{1}{c}{\makebox[-1in][r]{\bf (d)}}\\ [-0.3cm]
    \includegraphics[width=3cm]{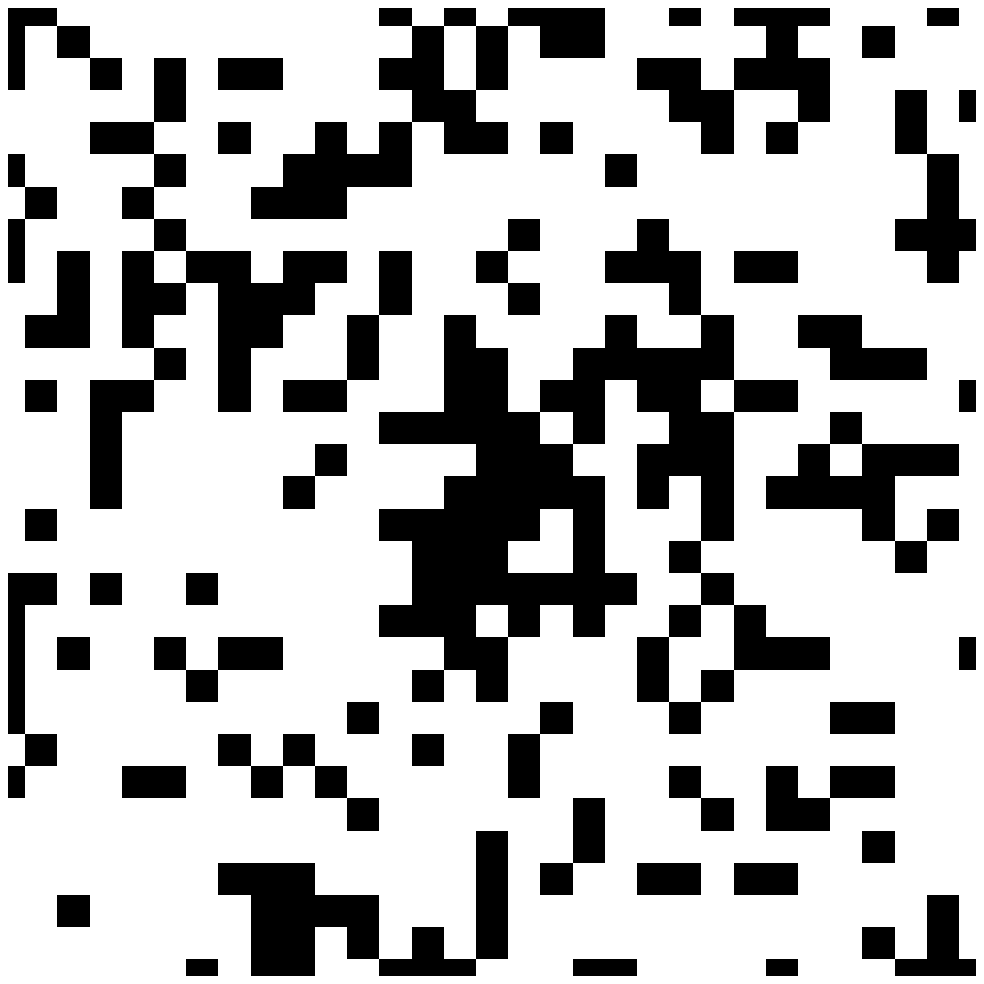}
    &\hspace{1.8cm} \includegraphics[width=4.5cm]{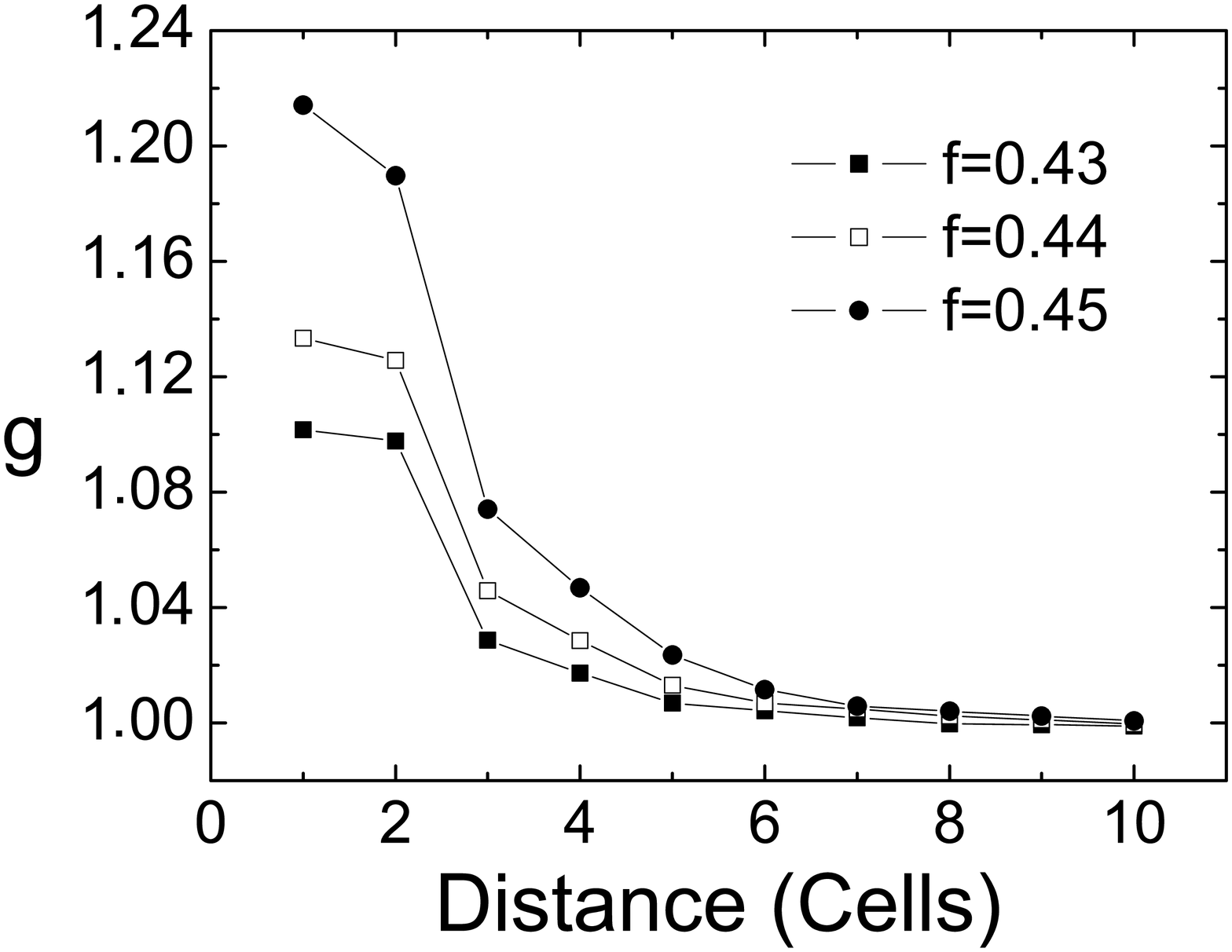}
  \end{array}$
   \caption{Savanna configuration in the clumped state at parameters as
   in Table \ref{parnamesComb2.2}. Panels (a) and (c) display only a $30\times30$ portion of the full
   $200\times200$ lattice. (a) $\delta= 0.01$, $f=0.31$: an example of open
   cluster of ATs, the typical configuration at these parameters. (b) Pair correlation function,
   similar to the one in the SM in the clumped state. (c) $\delta= 0.001$, $f=0.45$:
  an example of closed cluster of ATs, the typical configuration at these parameters. (d) Pair
  correlation function, which is different to the one in the SM in the clumped
  state because the maximum of $g(l)$ occurs at $l=1$, i.e. in the near neighborhood.
   }\label{opencluster}
\end{figure}

\begin{figure}[ht]
 \centering
 $\begin{array}{cc}
 \multicolumn{1}{c}{\makebox[-2.5in][r]{\bf (a)}}
 &\multicolumn{1}{c}{\makebox[-2.2in][r]{\bf (b)}}\\ [-0.9cm]
   \includegraphics[width=7cm]{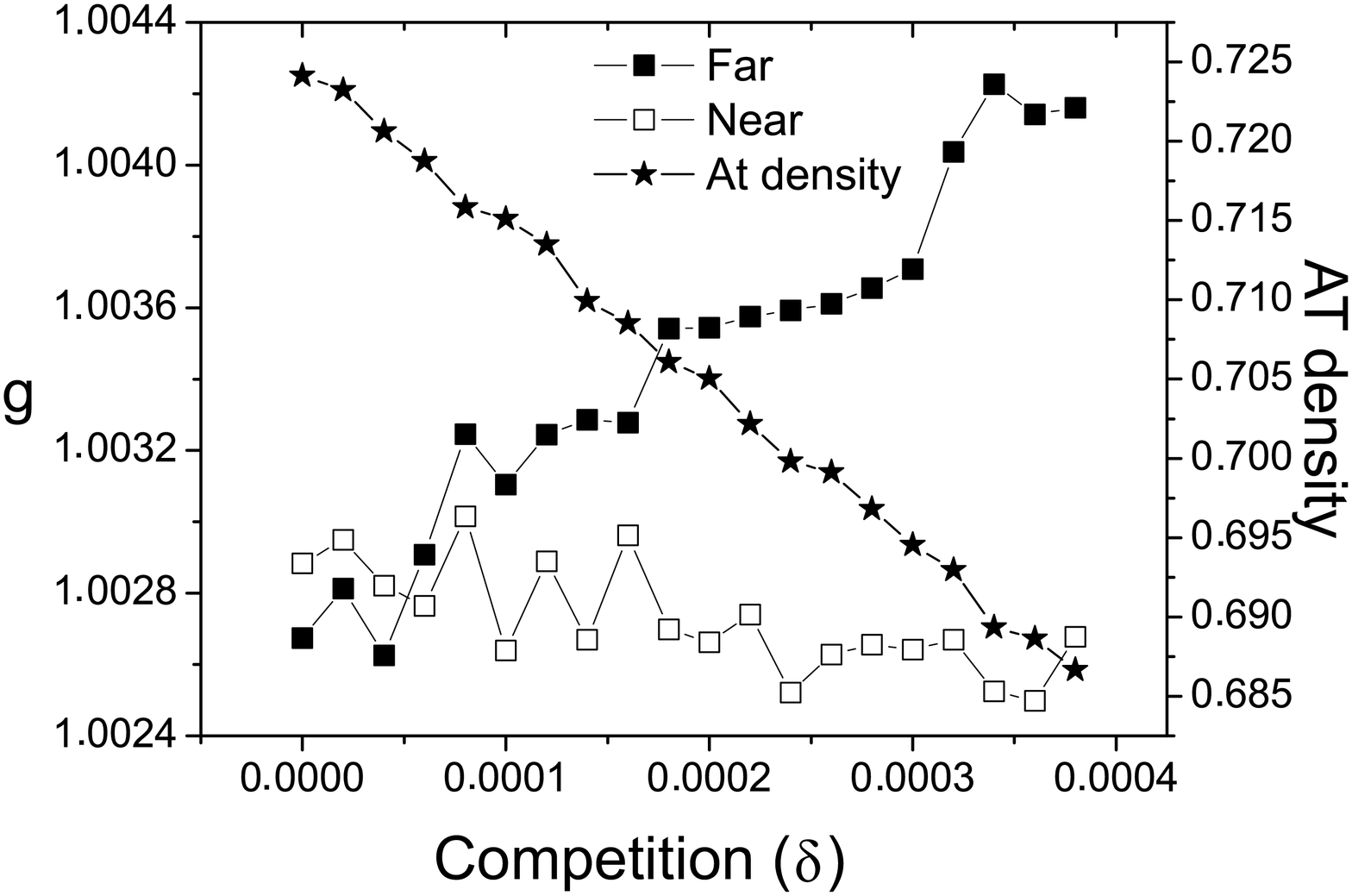} &
   \includegraphics[width=6 cm]{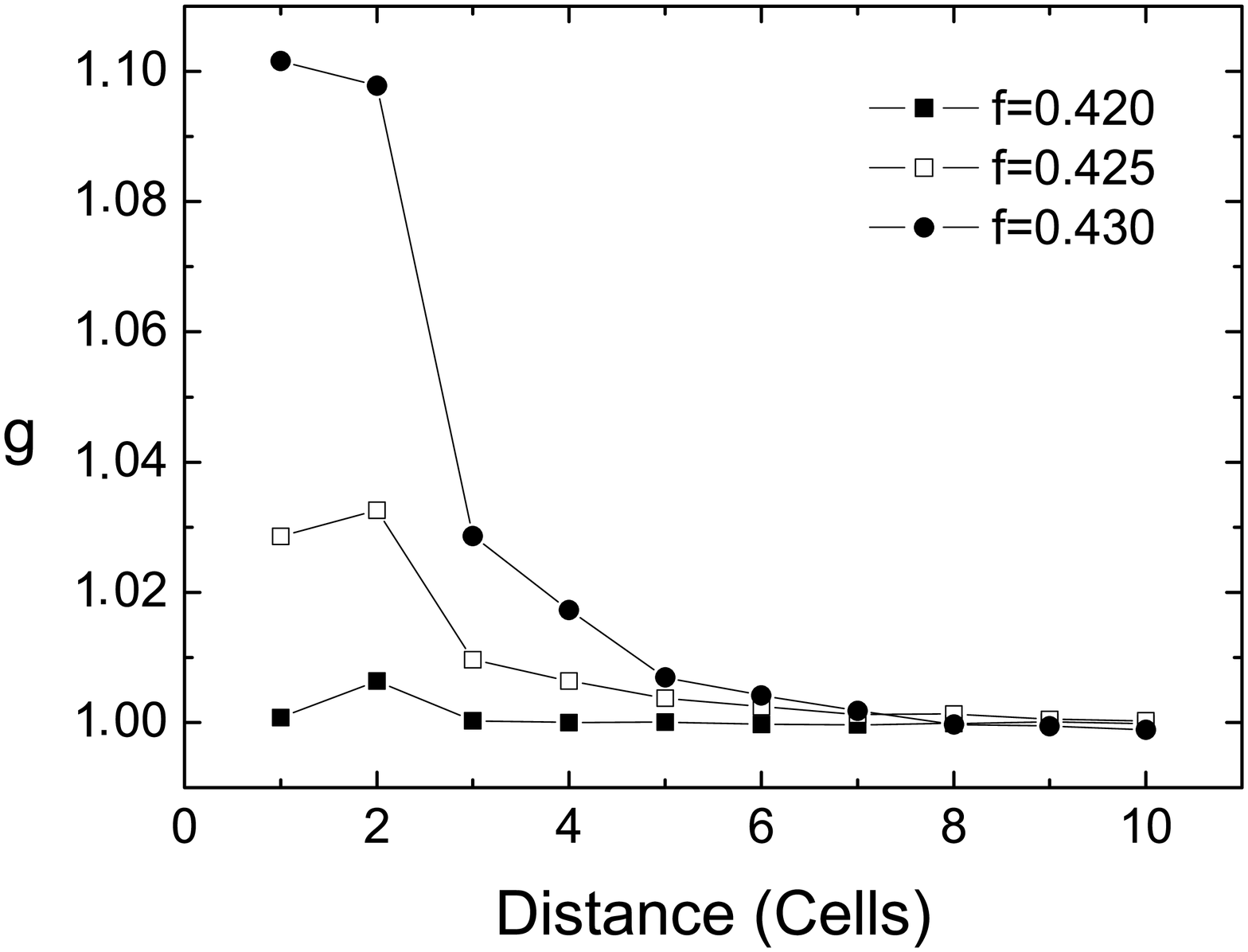}
 \end{array}$
    \caption{(a) Values of the pair correlation function for near neighbor pairs
   $g(1)$ and for far neighbors $g(2)$, and values of adult-tree density, both as a function
   of the tree-competition
   coefficient $\delta$. Closed clusters occur for $\delta<5.2 \cdot10^{-5}$
   and open ones for $\delta>5.2 \cdot10^{-5}$. (b) $\delta=0.001$, Pair correlation
   functions showing the transition between open ($g(1)<g(2)$) and closed ($g(1)>g(2)$) clusters by
   increasing $f$. Other parameters in both panels as in Table \ref{parnamesComb2.2}. }
   \label{gdeltainversion}
\end{figure}

In the clumped patterns just described, further illustrated by
Fig. \ref{opencluster}(a) and (b), the clusters are \emph{open}
in the sense that there are more neighbors in the far
neighborhood than in the near neighborhood. This a clear effect
of the competition existing in the near neighborhood, and was
the only clumped state present in the previous SM. The
novelty here is that, in addition, there is a second type of
clustered state not present in the SM. A clumped state
made of \emph{closed} clusters is illustrated by figure
\ref{opencluster}(c) and (d). The clusters are closed in the
sense that there are more AT neighbors in the near neighborhood
than in the far neighborhood. Thus, the positive effect of fire
protection (and local dispersal) has completely overcome the
competition effect occurring in the near neighborhood. The
transition from one type of pattern to the other occurs when
changing the competition or the lightning parameters, $\delta$
and $f$, as shown in Figure \ref{gdeltainversion}.

\vspace{5cm}
\subsection{Cluster-size distributions}

A cluster is a group of neighboring sites occupied by the same
type of vegetation (e.g. adult trees). The distribution of
cluster sizes is a powerful indicator of the different
mechanisms occurring in ecosystems
\citep{Pascual2002,Pascual2005}. Adult tree cluster-size
distributions in the Kalahari have been investigated by
\citet{scanlon2007}, finding that in most cases a power-law fit
can describe the data (although the fit was not of uniform
quality). \citet{scanlon2007} showed that resource constraints,
together with positive local interactions of the type
identified in the previous section, could generate cluster-size
distributions similar to the observed ones.

\begin{figure}
  \centering
  $\begin{array}{cc}
    \includegraphics[width=6cm]{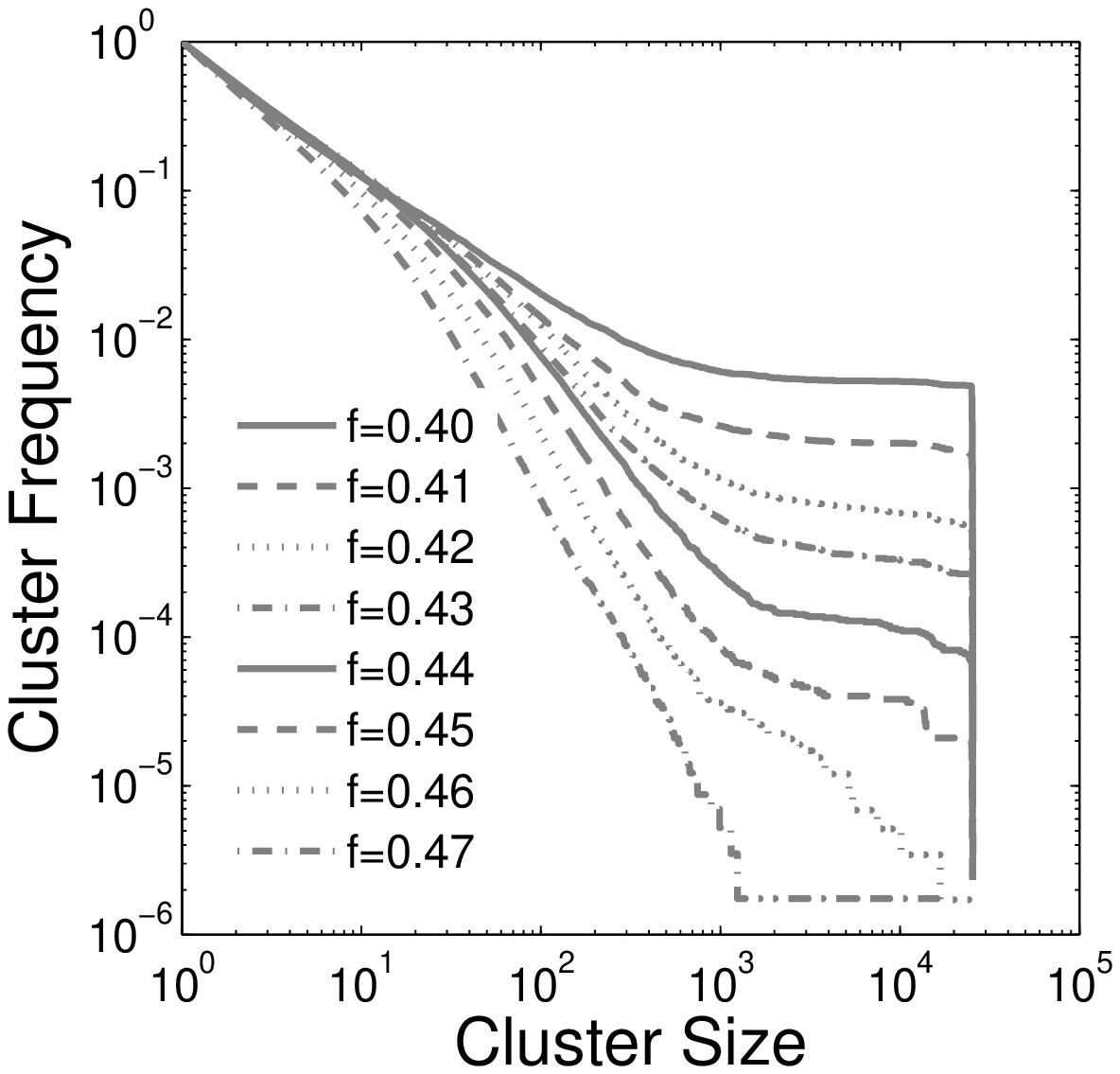}& \includegraphics[width=6cm]{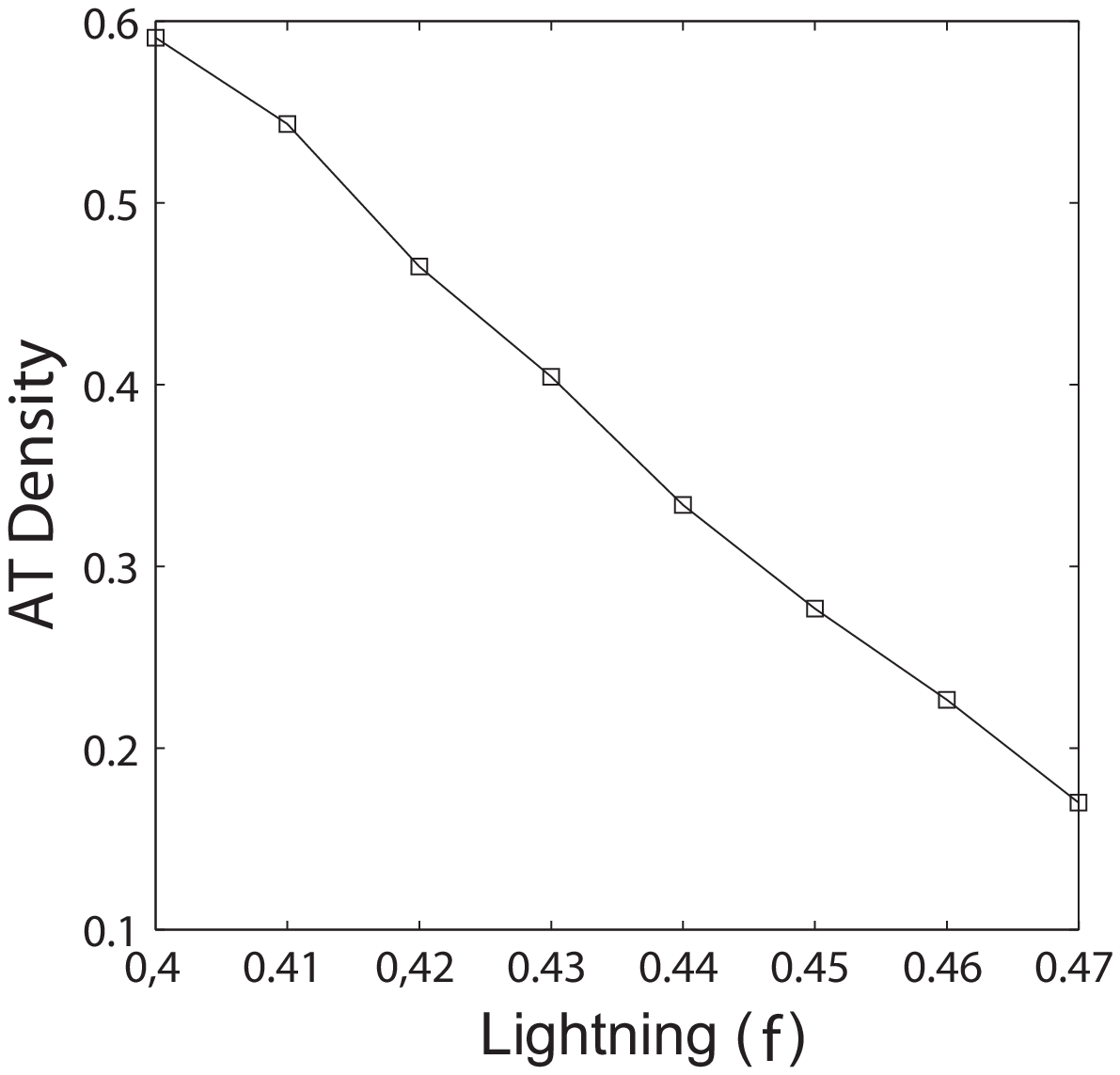}
  \end{array}$
  \caption{Left: Adult-tree cluster-size distributions, represented by means of the
   complementary cumulative distributions.
  Parameters as in Table \ref{parnamesComb2.2} but $\delta$= 0.001. Right: Density of adult trees
  (i.e., the number of adult trees divided by the total number
of lattice sites) versus lightning $f$.
}
  \label{cfdCLUSTER}
\end{figure}

Figure \ref{cfdCLUSTER} shows complementary cumulative
distributions of adult-tree cluster sizes from our model, where
the Moore neighborhood has been used to define clusters. Though
the distributions have fat tails (see Fig. \ref{cfdCLUSTER}), a
single power law does not provide a good description in the
realistic range of parameters considered above. Also, the
plateau at large sizes in the small-$f$ curves of Fig.
\ref{cfdCLUSTER} indicate the presence of clusters much larger
than the rest. By artificially changing parameters to other
ranges, one can find situations in which the cluster-size
distribution follows a relatively good power law. This happens,
for example, for $\alpha=\beta=1$, $\delta=0.01$, and $f\approx
0.9$.
Inspection of the tree distributions above and below this
$f$-value indicates that a percolation transition occurs
precisely at that point: there is a giant AT cluster spanning
the whole area for smaller $f$ values, and disconnected tree
patches for higher values. Power-law cluster distributions are
observed then close to this percolation transition, as in the
mechanism discussed in \citet{Pascual2005}, although in a
narrower parameter range than suggested there. This transition
is not attained within the parameter ranges considered before
in this paper: tree cover in Fig. \ref{cfdCLUSTER} is just of
0.4 for $f\approx 0.9$, and can not be increased much more (see
Fig. \ref{p-density}, right panel), which makes it difficult to
attain percolation through the whole lattice because of the
absence of very large clusters. By artificially changing
parameters to obtain larger tree densities, percolation becomes
easier. In such situations, we observe more robust power-law
behavior (not shown), but the system is then closer to a forest
than to a realistic savanna. We do not find systematic
correlation between the small-scale character of the tree
patterns (regular, clumped, open, closed, ...) and the type of
cluster-size distributions, despite the fact that one could
expect that positive short-range correlations would favor
power-laws \citep{scanlon2007}.

\section{Summary}

We have introduced a model for savanna structure which
includes, in addition to fundamental ecological interactions
including competition, the effects of spatially explicit fire
spread. Fire introduces some effective tree-grass and tree-tree
interactions which are important in shaping tree demography and
spatial pattern. First, the presence of fire improves
competitiveness of grass because of its faster post-fire
recovery. Second, adult trees may protect nearby juveniles from
fire. This results in a positive tree-tree interaction which
can, in some circumstances, overcome the effects of tree-tree
competition for resources. A variety of tree spatial
distributions are observed as a result of these direct and
indirect interactions, which we have characterized by the pair
correlation function and the cluster-size distribution. As the
short-range positive interactions gain importance relative to
the negative ones, a succession of regular to clumped states is
observed. Clumped states can have ``open'' clusters, like the
ones present in the previous SF model \citep{Calabrese2010},
but also ``closed'' clusters for the cases with stronger
positive interactions. Adult-tree cluster-size distributions
are of power-law type in some cases because of the proximity to
a percolation transition, but for much of the realistic
parameter range tree cover is small and far from percolating.
The tails of the distributions, although fat, seem to decay
faster than power laws, as seen in fact to occur in several of
the sites reported by \citet{scanlon2007}.

\section*{Acknowledgment}
F.S.B. and E.H.-G. acknowledge financial support from Spanish
MINECO and FEDER through project FISICOS (FIS2007-60327). J.M.C
acknowledges the support of the European Union project PATRES
(Pattern Resilience; project NEST 43268). F.S.B. acknowledges a
grant from the Balearic Government. We thank Volker Grimm and
Richard D. Zinck for helpful discussions.


\clearpage 


%
\clearpage    
\section*{Figures}

\end{document}